\documentstyle[12pt]{article}

\newcommand{\sect}[1]{\setcounter{equation}{0}\section{#1}}
\newcommand{\subsect}[1]{\subsection{#1}}
\newcommand{\subsubsect}[1]{\subsubsection{#1}}

\newcommand{\lb}[1]{\label{eq:#1}}
\newcommand{\rf}[1]{(\ref{eq:#1})}

\newcommand{\ra}{\rightarrow}

\newcommand{\lra}{\longrightarrow}

\def\col#1,#2,#3,#4 {\begin{array}{c}#1\\ #2\\ #3\\#4 \end{array}}
\def\prop #1.{\bigskip \noindent {\bf Proposition #1.} \par}
\def\proof{\bigskip \noindent {\it Proof.} \ }
\def\d#1,#2{\frac{d#1}{d#2}}
\def\pd#1,#2{\frac{\partial#1}{\partial#2}}
\def\lpd#1,#2{\frac{\stackrel{\rightarrow}{\partial#1}}{\partial#2}}
\def\rpd#1,#2{\frac{\stackrel{\leftarrow}{\partial#1}}{\partial#2}}
\def\td#1{\dot{#1}}
\def\ttd#1{\ddot{#1}}
\def\tev{\frac{d}{dt}}
\def\fraz#1,#2{\frac{#1}{#2}}

\def\vq{\vec{q}}

\def\tvq{\td{\vec{q}}}
\def\bq{{\bf q}}
\def\tbq{\td{\bf q}}
\def\vp{\vec{p}}
\def\wt{\widetilde}
\def\inner{\underline {\ \ }\kern -.1em \raise.3ex\hbox{$\vert $}\,}

\def\bsk{\bigskip}
\def\be{\begin{equation}}
\def\ee{\end{equation}}
\def\bea{\begin{eqnarray}}
\def\eea{\end{eqnarray}}
\def\nn{\nonumber}
\def\fl{\forall~}
\def\om{\omega}
\def\de{\delta}

\def\si{\sigma}
\def\th{\theta}
\def\wth{\wt{\th}}
\def\ta{\tau}

\def\Ga{\Gamma}
\def\La{\Lambda}
\def\Si{\Sigma}

\def\thl{\th_{\L}}
\def\oml{\om_{\L}}
\def\fkg{\bf g}
\def\mcl{g^{-1}dg}
\def\mcr{(dg)g^{-1}}
\def\mcdl{g^{-1}\td{g}}
\def\mcdr{\td{g} g^{-1}}
\def\me{{\sf m}}
\def\A{{\cal A}}

\def\D{{\cal D}}
\def\F{{\cal F}}

\def\I{{\cal I}}
\def\L{{\cal L}}

\def\N{{\cal N}}

\def\P{{\cal P}}
\def\Q{{\cal Q}}
\def\R{{\cal R}}

\def\X{{\cal X}}

\def\Cb{{\bf C}}
\def\Hb{{\bf H}}

\def\Rb{{\bf R}}

\def\Gb{{\bf G}}
\def\Mb{{\bf M}}
\def\Qb{{\bf Q}}
\def\Xb{{\bf X}}
\def\func{\F(M)}
\def\vect{\X(M)}

\def\tanb{TQ}
\def\cotb{T^*Q}

\def\vectq{\X(\tanb)}

\def\funcc{\F(\cotb)}

\begin{document}

\thispagestyle{empty}

\hfill Wien, 9 June 93

\hfill ESI 28 (1993)

\vspace{.5cm}

\begin{center}
{\LARGE \bf Generalized Reduction Procedure:
\\~\\
Symplectic and Poisson Formalism}
\end{center}

\vspace{.5cm}

\begin{center}
{\Large J. Grabowski},
{\Large G. Landi}\footnote{Partially supported by the Italian
Consiglio Nazionale delle Ricerche
under Grant 203.01.61.}, {\Large G. Marmo and G. Vilasi}
\end{center}

\bigskip

\begin{center}
{\it E. Schr\"odinger International Institute for Mathematical Physics, \\
Pasteurgasse 4/7, A-1090 Wien, Austria.}
\end{center}

\vspace{2.0cm}

\begin{abstract}
We present a generalized reduction procedure which encompasses the
one based on  the momentum map and the projection method. By using
the duality between manifolds and ring of functions defined on them,
we have cast our procedure in an algebraic context. In this framework
we give a simple example of reduction in the non-commutative setting.
\vspace{1.0cm}
\end{abstract}

\vfill\eject

\thispagestyle{empty}

\noindent Permanent addresses.

\bigskip

\begin{center}

{\Large J. Grabowski}\\
Institute of Matematics - University of Warsaw,\\
ul. Banacha 2, PL 02-097 Warsaw --- Poland.\\
jagrab@mimuw.edu.pl

\bigskip

{\Large G. Landi}\\
Dipartimento di Scienze Matematiche - Universit\`a di Trieste,\\
P.le Europa, 1 - I-34100 Trieste --- Italy.\\
landi@univ.trieste.it\\
and\\

\smallskip
Istituto Nazionale di Fisica Nucleare - Sezione di Napoli\\
Mostra d'Oltremare, Pad.20 - I-80125 Napoli --- Italy.

\bigskip

{\Large G. Marmo}\\
Dipartimento di Scienze Fisiche - Universit\`a di Napoli,\\
Mostra d'Oltremare, Pad.19 - I-80125 Napoli --- Italy.\\
gimarmo@na.infn.it\\
and\\

\smallskip
Istituto Nazionale di Fisica Nucleare - Sezione di Napoli\\
Mostra d'Oltremare, Pad.20 - I-80125 Napoli --- Italy.

\bigskip

{\Large G. Vilasi}\\
Dipartimento di Fisica Teorica e S.M.S.A. - Universit\`a di Salerno,\\
Via S. Allende, I-84081 Baronissi (SA) --- Italy.\\
vilasi@sa.infn.it\\
and\\

\smallskip
Istituto Nazionale di Fisica Nucleare - Sezione di Napoli\\
Mostra d'Oltremare, Pad.20 - I-80125 Napoli --- Italy.
\end{center}

\pagebreak

\noindent
\sect{Introduction}

It is well known that a constant of the motion can be used to ``reduce the
number of degrees of freedom" of a Hamiltonian dynamical system [Wi],
[AM], [Ar], [MSSV].
An adequate number of constants of the motion may give rise to
{\it action-angle}
variables which are then used to analyze completely integrable systems.

In the Hamiltonian or Lagrangian formalism,
constants of the motion are
associated with symmetries of the equations of the motion and a
reduction procedure relies on both concepts.
First of all, constants of the motion provide
invariant submanifolds, while symmetries, when acting on them, provide
equivalence classes. The reduced dynamics becomes a dynamics on the
quotient manifold of equivalence classes. Very often the reduced space
can be imbedded as an invariant
submanifold of the starting carrier space and one gets
a dynamical evolution of initial data which have already a
physical interpretation.

For instance, in the Hamiltonian formalism,
a group $G$ acts symplectically on the symplectic manifold
$(M,\omega )$ and defines a momentum map $\mu : M \to \fkg^*$.
This map gives the Hamiltonian generators of the action
of $G$ by setting $i_{X_j}\omega
= d \mu (e_j)$, where $e_1, e_2,\ldots , e_r$, is a basis for
the Lie algebra $\fkg$ of $G$, and $\fkg^*$ is the dual vector space.
Given the regular value $k \in \fkg^*$ of $\mu$, one considers the
submanifold $\Sigma_k= \mu ^{-1} (k)\subset M$
on which $G_k$, the stability group of $k$ under the
coadjoint representation, acts.
The quotient manifold $\wt\Sigma _k = \Sigma_k / G_k$ is
symplectic and is called the reduced phase space.
If $G$ is a symmetry group for the original dynamics, the latter
preserves $\Sigma _k$ and
projects onto a Hamiltonian dynamics on $\wt\Sigma _k$.
In this construction, the group $G$ provides both the
submanifold $\Sigma_k$ and the subgroup $G_k$ used
to foliate it in order to get $\wt\Sigma _k$.

An interesting feature of the reduced system is that it may turn
out to be non linear even if the starting one was linear and therefore
integrable via exponentiation. Many completely integrable systems actually
turn out to be reduced systems of free or simple ones [OP].
One may be tempted
to conjecture that any completely integrable system should arise as reduction
of a simple one. By simple here we mean a system which can be
integrated via exponentiation (i.e. like going from a Lie algebra to a
corresponding Lie group).
It seems therefore
convenient to elaborate a procedure which would be a kind of converse
of the reduction procedure  [KKS].
One hopes to unfold the non linearity
of the dynamics on the quotient manifold  by going to a higher dimensional
carrier space so to get a system whose flow can be easily found.
By applying the  reduction procedure one then provides
a flow for the starting non linear system.
Of course there is no prescription to unfold nonlinearities, therefore a
better understanding of the reduction procedure
might help in suggesting a converse procedure.

Various descriptions of dynamical
systems are available. One starts with a vector field $\Gamma$ on a
manifold $M$ (carrier space) and by further qualification of $M$ and
$\Gamma$ one finds  a Lagrangian description on $M=TQ$ and $\Gamma$
second order [AM], [MSSV], [MFLMR]
or a symplectic description on (M, $\omega$), with
$\omega$ a symplectic structure and $\Gamma$ a symplectic vector field.
When (M, $\Gamma$) is thought of as a classical limit of a quantum system,
$M$ will be required to be a Poisson manifold and $\Gamma$
to be a Hamiltonian
vector field with respect to the given Poisson structure.

When dealing with the reduction of systems using additional structures
one is obliged to worry also about them, i.e. the reduction procedure
should be
compatible with the additional structure on the carrier space.
In some previous work we have considered some examples with the
generalized reduction at work. We did not, however, consider the
r\^ole of additional structures like the symplectic or the Poisson
structure. Here we would like to take into account these additional
structures. We shall also consider few examples where the Poisson
structures are of the Lie-Poisson type and an example in the
non-commutative (quantum) geometry.

As for notations we shall refer to [AM], [LM], [MSSV].
To help visualize how the paper  is organized, we give a list of
contents:
\begin{description}
\item[1] Introduction.
\item[2] Generalized Reduction Procedure.
Examples: rotationally invariant dynamics.
Reduction and symplectic structures.
Reduction of free motion and symplectic description.
\item[3] Reducing geodesical motion on Lie groups.
Few elements of Lagrangian formalism.
Geodesical motion on Lie groups.
Geodesical motion on $SU(2)$.
\item[4] The Calogero-Moser system.
{}~Symplectic ~reduction ~and ~deformation ~for ~the ~
         Calogero-Moser system.
\item[5] Reduction of geodesical motions.
Geodesical motions on spheres.
Geodesical motions on hyperboloids.
\item[6] Reduction procedure in algebraic terms.
Poisson reduction.
\item[7] An example of non-commutative reduction.
\item[8] Conclusions.
\end{description}

\bigskip

\bigskip

\noindent
\sect{Generalized Reduction Procedure}
Let us consider a dynamical system $\Gamma$ on a carrier space $M$,
namely an element $\Ga \in \vect$. We shall indicate by $\Phi_t^{\Ga}$ its
flow.

A generalized reduction procedure is based on two steps [LMSV], [MM].

\noindent
In the first one we consider:
\begin{description}
\item[i)]
A submanifold $\Si$ of $M$ which is invariant for $\Ga$, namely
to which $\Gamma$ is tangent,
\be
\Gamma(m) \in T_m\Si~,~~~\fl~ m \in\Si~.
\ee
\item[~]
In the second one we consider:
\item[ii)]
An equivalence relation ${\cal R}$ on $\Si$ which is compatible with
$\Gamma$, namely such that
\be
m~ {\cal R}~ m' \iff \Phi_t^\Gamma(m)~ {\cal R}~ \Phi_t^\Gamma(m')~,
{}~~~\fl~ m, m' \in \Si~.
\ee
\end{description}

\noindent
Given the previous ingredients, the {\it reduced} carrier space is the
set of equivalences classes
\be
\wt \Si = \Si / \cal R~,
\ee
and the reduced dynamical system $\wt \Gamma$ will be
the projection of $\Gamma$ along the natural projection
$\pi_{\cal R}: \Si\to \wt\Si_{\cal R}$.

\bsk

\noindent {\bf Remark.}
Either one of the two previous steps
could be trivial. Moreover, one could also proceede in
opposite order, by giving first a compatible equivalence
relation on $M$ and then selecting an invariant submanifold in
$M / \R$.

\bsk

As we shall see, there are several ways to get an equivalence relation
on invariant submanifolds. The most common ones are gotten by means of the
action of a group or by involutive distributions.

Suppose we have a Lie group $G$
which acts on $\Sigma$ and which is a
symmetry group for $\Gamma$ restricted
to $\Sigma$, (we do not require $G$ to act on the full
manifold $M$, and if it acts on $M$ we do not require
it to be a symmetry group for $\Gamma$ on $M$ but only
for the restriction to $\Sigma$). On the quotient
$\wt{\Sigma}=\Sigma/G$ we get a reduced dynamics.

\bsk

We can `dualize' the previous scheme by giving it on
the algebra $\func$ of observables on $M$.
This dual view point is
usefull e. g. when dealing with Poisson dynamics.
In this algebraic context a dynamical system is seen as
an element $\Ga \in Der \F$ giving
\be
\dot f = \Gamma \cdot f, ~~~\fl~ f\in {\cal F}~.
\ee

A reduction procedure will again require two steps.

\begin{description}
\item[ia)] There exists an algebra $\F_\Si$ and a projection
\be
\pi_\Si : \F \to \F_\Si~ \lb{redAA}
\ee
(this projection can be visualized in terms of the identification map
$i_\Si:\Si \to M$ by setting $\pi_\Si(f) = i^*_\Si(f)$,
so that $\F_\Si$ plays the r\^ole of the algebra of functions on
$\Si$),
\item[~]
and a derivation $\Ga_\Si \subset Der \F_\Si$ such that
\be
\pi_\Si (\Gamma \cdot f) = \Gamma_\Si \cdot \pi_\Si (f)~~~\fl f
\in \F
\ee
(this translates the invariance condition on $\Si$).
\item[iia)]
There exists an invariant subalgebra of $\F_\Si$, namely
a subalgebra
$\wt {\cal F} \in {\cal F}_\Si$ such that
\be
{\Gamma }_\Si \cdot \wt {\cal F} \subset \wt {\cal F}
\ee
(this translates the compatibility condition).
\end{description}

The restriction of $\Gamma_\Si$ to $\wt {\cal F}$ can be
denoted by $\wt \Gamma$ and provides us with the analog of the reduced
dynamics we had an $\wt\Si$.

\bsk

We can visualize our dual two steps reduction procedures with the
help of the
following diagrams
\be \setlength{\unitlength}{1mm}
\begin{picture}(110,40)(-14,7)
\put(0,25){$(\Si ,\Gamma_\Si)$ }
\put(15,25){\vector(2,-1){16}}
\put(15,27){\vector(2,1){16}}
\put(32,15){$(\wt\Si,\wt\Gamma)$}
\put(32,35){$(M,\Gamma)$}
\put(38,32){\vector(0,-1){12}}
\put(20,33){$i_{\Si}$}
\put(20,18){$\pi_\R$}

\put(65,15){$(\wt{\cal F},\wt\Gamma)$}
\put(65,35){$({\cal F},\Gamma)$}
\put(70,20){\vector(0,1){12}}
\put(77,35){\vector(2,-1){16}}
\put(77,17){\vector(2,1){16}}
\put(84,32){$\pi_\Si$}
\put(84,17){$i_\R$}
\put(95,25){$({\cal F}_\Si,\Gamma_\Si)$}
\end{picture}
\ee
where
$
i_{\cal R}(\wt{\cal F})\ =\ \pi_{\cal R}^{\ast }({\cal F} (\wt \Si ))
$
and
$
\pi _{\Si }({\cal F})=i_{\Si }^{\ast }({\cal F}).
$

\bsk

\noindent {\bf Remark.}
Here we should point out that if $i_\Si:\Si \to M$ is an embedding, then
$\F_\Si \equiv \F(\Si)$, while if $i_\Si$ is only an immersion, then
$\F_\Si \subset \F(\Si)$.

\bigskip

Before we move on to take into account additional structures we
illustrate the reduction procedure by using the ring of
functions in the following
examples.

\bigskip

\subsect{Examples: rotationally invariant dynamics}
In these examples we shall consider the less familiar
dual point of view.

Suppose we have a dynamical system $\Ga$ on $M$ which is invariant
under the action of a Lie group $G$, namely $G_* \Ga = \Ga$.
To construct a reduced dynamics we consider ${\cal F}^G
= \{ f\in {\cal F} (M) \ ,\ G^*\cdot f=f\}$, i.e.
the algebra of invariant functions under the action of $G$.
Under some regularity assumptions on the action of $G$, the
real spectrum ${\cal N}_G$
of ${\cal F}^G$ determines a differential manifold on which $\Gamma$ defines
a dynamical vector field $\Gamma^*$. Any ideal of constants of
the motion for $\Gamma^*$ in ${\cal F}^G$ gives a quotient
algebra on which $\Gamma^*$ defines a reduced dynamics
$\wt\Ga$. For details on these algebraic statements see section
\ref{se:red}.

As an example let us consider a
dynamical system $\Ga$ on $T\Rb^3$, invariant
under the action of the rotation group $SO(3)$.
If we suppose that $\Ga$ is second order, its general expression
will be
\be
\Ga = \td{\vec r} \cdot \pd{},{\vec r} + \vec f ( r, \td{r},
\vec r \cdot \td{\vec r}) \cdot \pd{},{\td{\vec r}}~, \lb{ex1}
\ee
and for the moment we do not make any assumption on the explicit form
of $\vec f$.
The algebra $\F^G$ is made of functions which are invariant under
$SO(3)$.
The space of the orbits
${\cal N}_G$ can be parametrized  by the three functions
$\xi_1=\vec r\cdot\vec r$,
{}~$\xi_2= \td{\vec r} \cdot \td{\vec r}$~ and
{}~$\xi_3= \vec r \cdot \td{\vec r}$~ which
are rotationally
invariant.
The reduced dynamics is then given by
\bea
&& {d\over dt}\xi_1 = 2 \dot{\vec r} \cdot \vec r = 2\xi_3~, \nn \\
&& {d\over dt}\xi_2 = 2 \dot{\vec r} \cdot \ddot{\vec r} =
2 \dot{\vec r} \cdot \vec f~, \nn \\
&& {d\over dt}\xi_3 = \dot{\vec r} \cdot \dot{\vec r} +
\vec r \cdot \ddot{\vec r} = \xi_2 + \vec r \cdot \vec f~, \lb{ex2}
\eea
and the assertion that ${d\over dt} \xi _j$ can be expressed
in terms of the $\xi$'s follows from the assumption of rotational
invariance for the starting dynamics.

If one wants to produce reduced dynamics with additional structures
(e.g. second order)
one needs an extra step. One has to select a subset of variables (in
this case just two) and express the dynamics in terms of them and of
constants of the motion.
To show this we shall specify our system as a very simple one,
namely the free particles.
In this case we have,
\bea
&&{d\over dt} \xi_1 = 2 \xi_3~, \nn \\
&&{d\over dt} \xi_2 = 0~, \nn \\
&&{d\over dt} \xi_3 = \xi_2~. \lb{ex3}
\eea

Now we select an invariant two-dimensional
submanifold in ${\cal N}_G$. This can be done in different ways.

\bigskip

\noindent
1. Fix $\xi_2 = k = const$. On this invariant submanifold the reduced
dynamics is
\bea
&&{d \over dt} \xi_1 = 2\xi_3~, \nn \\
&&{d \over dt} \xi_3 = k~. \lb{ex4}
\eea
If we set $x = \xi_1$ and $v = 2 \xi_3$, the resulting
system is a Lagrangian one with Lagrangian function
given by
\be
\L = {1\over 2} v^2 + 2k x~. \lb{ex5}
\ee

\bigskip

\noindent
2. Fix $\xi_1\xi_2 - \xi^2_3 = \ell^2 = const$ (this is the square of
the angular momentum). We get
$\xi_2 = {1\over \xi_1} (\xi^2_3+\ell^2)$ and as reduced dynamics
\bea
&&{d\over dt}\xi_1 = 2\xi_3~, \nn \\
&&{d\over dt} \xi_3 = {\xi^2_3 + \ell^2 \over \xi_1}~. \lb{ex6}
\eea
If we set again $x = \xi_1$ and $v = 2 \xi_3$, the resulting
system is a Lagrangian one with Lagrangian function
given by
\be
\L = {1\over 2} {v^2 \over x} - {2\ell^2 \over x}~. \lb{ex7}
\ee

\bigskip

Let us consider now what happens if we try to get a dynamics on a
different set of (invariant) variables while mantaining the same
invariant submanifold. Suppose we take as variables
$\eta = r = \sqrt{\xi_1}$ and $\td{\eta}$. Starting from the free
motion, after some algebra we find
\be
\tev \td{\eta} = {L^2 \over \eta^3}~, \lb{ex8}
\ee
where $L^2$ is the square of the angular momentum.
If we now fix $L^2 = \ell^2$ we get a reduced system of Calogero-Moser
type
\be
\tev \td{\eta} = {\ell^2 \over \eta^3}~, \lb{ex9}
\ee
which is quite different from the \rf{ex6}. The corresponding
Lagrangian function is given by
\be
\L = {1\over2} (\td{\eta}^2 - {\ell^2 \over \eta^2})~. \lb{ex10}
\ee

\bigskip

We can do similar considerations in the Hamiltonian framework.
As an example let us consider now a
dynamical system $\Ga$ on $T^*\Rb^3$, invariant
under the action of the rotation group $SO(3)$.
If we suppose that $\Ga$ is Hamiltonian with Hamiltonian function
$H$ invariant under the canonical action of $SO(3)$,
its general expression,
with respect to the standard symplectic structure, will be
\be
\Ga = \pd{H},{\vec p}(r, p , \vec r \cdot \vec p) \cdot \pd{},{\vec r}
- \pd{H},{\vec r} (r, p, \vec r \cdot \vec p) \cdot \pd{},{\vec p}~.
\lb{ex11}
\ee
The space of orbits
${\cal N}_G$ can now be parametrized  by the three functions
$\xi_1=\vec r\cdot\vec r$,
$\xi_2= \vec p \cdot \vec p$ and
$\xi_3= \vec r \cdot \vec p$.
The reduced dynamics is then given by
\bea
&& {d\over dt}\xi_1 = 2 \dot{\vec r} \cdot \vec r
= 2 \vec r \cdot \pd{H},{\vec p}~, \nn \\
&& {d\over dt}\xi_2 = 2 \vec p \cdot \dot{\vec p} =
- 2 \vec p \cdot \pd{H},{\vec r}~, \nn \\
&& {d\over dt}\xi_3 = \dot{\vec r} \cdot \vec p +
\vec r \cdot \dot{\vec p} = \vec p \cdot \pd{H},{\vec p}
- \vec r \cdot \pd{H},{\vec r}~, \lb{ex11a}
\eea
and the assertion that ${d\over dt} \xi _j$ can be expressed
in terms of the $\xi$'s follows from the assumption of rotational
invariance for the starting Hamiltonian.

We take again the free particle. Then the reduced system is formally
the same as in \rf{ex3},
\bea
&&{d\over dt} \xi_1 = 2 \xi_3~, \nn \\
&&{d\over dt} \xi_2 = 0~, \nn \\
&&{d\over dt} \xi_3 = \xi_2~. \lb{ex12}
\eea

In order to have a reduced dynamics which is symplectic, we have to
select an invariant two-dimensional
submanifold in ${\cal N}_G$.

\bigskip

\noindent
1. Fix $\xi_2 = k = const$. On this invariant submanifold the reduced
dynamics is
\bea
&&{d \over dt} \xi_1 = 2\xi_3~, \nn \\
&&{d \over dt} \xi_3 = k~. \lb{ex13}
\eea
This system is Hamiltonian with respect to the symplectic structure
\be
\omega =d\xi_1\wedge d \xi_3~,~~~~~
H= \xi^2_3 - k\xi_1 ~. \lb{ex14}
\ee

\bigskip

\noindent
2. Fix $\xi_1\xi_2 - \xi^2_3 = \ell^2 = const$. As reduced dynamics we
get
\bea
&&{d\over dt}\xi_1 = 2\xi_3~, \nn \\
&&{d\over dt} \xi_3 = {\xi^2_3 + \ell^2 \over \xi_1}~. \lb{ex15}
\eea
This is Hamiltonian with respect to the symplectic structure
\be
\omega = {1\over \xi^2_3 +\ell ^2}\  d\xi_1\wedge d\xi_3~,~~~~~
H={1\over m} ln {\xi^2_3 +\ell^2\over |\xi_1|} ~. \lb{ex16}
\ee

\bigskip

The examples we have considered have shown that even from a simple
system like the free particle, by using our reduction procedure it is
possible to obtain a variety of interacting systems
all of them however, completely integrable ones.
\bigskip

\bigskip

\subsect{Reduction and symplectic structures}
In the previous example we have already seen how one may get reduced
Hamiltonian systems starting with a Hamiltonian one.

Here we make some general considerations. We have
already emphasized that eventually we would like to reduce simple
systems to get completely integrable ones. We would
like to consider for instance, the reduction of free systems or harmonic
oscillators. It is known [MSSV] that these systems admits many
alternative Lagrangian or Hamiltonian descriptions. Therefore, it is
interesting to investigate if these alternative descriptions survive
the reduction procedure, i.e. do they provide alternative
descriptions for the reduced dynamics? We do know that in many cases
the answer is positive because many completely integrable systems do
possess alternative descriptions [DMSV]. Let us then examine the
situation.

We start with a symplectic manifold $(M, \om)$ and require that the
dynamics $\Ga$ is $\om$-Hamiltonian with Hamiltonian function $H$,
\be
i_\Gamma\om = dH~.
\ee

On any invariant submanifold
$ \Si ~{\buildrel i_\Si \over \hookrightarrow}~ M$
we get a 2-form by pulling back
$\om$, $\om_\Si = i_\Si ^*\om$ which, however,
will be degenerate in general.
We can consider now the equivalence relation associated with the
distribution defined by $ker\om_\Si$.
It is clear that if $i_{X}\om_\Si = 0$ for some vector field $X
\in \X(\Si)$, we find that $0 = L_{\Ga} i_{X}\om_\Si = i_{[\Ga,
X]}\om_\Si$. With some abuse of notation we do not distinguish $\Ga$
from $\Ga \vert_{\Si}$. Since $\Ga$ preserves the distribution
$ker \om_\Si$, it will be compatible with the associated equivalence
relation.

{}From the dual point of view, we can consider all Hamiltonian vector
fields on $M$ which are tangent to $\Si$, i.e. $i_{X_f}\om = df$ and
$X_f(m) \in T_m\Si~,~\forall m \in \Si$. The pull back to $\Si$ of
all these Hamiltonian functions will provide us with a subalgebra of
$\F(\Si)$ that will be denoted $\F(\om,\Si)$. The dynamical vector
field $\Ga$ on $\Si$ will map this subalgebra into itself; therefore,
its action on such a subalgebra provides the reduced dynamics
$\wt{\Ga}$. This statement follows easily from the fact that
$i_{X_f}\om = df$ can be restricted ``term by term" to $\Si$, because
$X_f$ is assumed to be tangent to $\Si$ and
$i_{X_f\vert_{\Si}}\om = df_{\Si}$.
{}From $L_{\Ga} i_{X_f}\om = i_{[\Ga, X_f]}\om = d( {\Ga}\cdot f)$, followed
by restriction to $\Si$ we prove our statement.

It is clear now that if we consider our reduced dynamics $\wt{\Ga}$
on the algebra $\F(\om,\Si)$, it is very easy to find out if another
symplectic structure $\om_1$ will be compatible with the reduction
procedure and provide an invariant two form for $\wt{\Ga}$. Indeed,
if we construct $\F(\om_1,\Si)$, the compatibility condition reads
$\F(\om_1,\Si) \subset \F(\om,\Si)$. Of course, if the two algebrae
coincide, $\F(\om_1,\Si) = \F(\om,\Si)$, $\om_1$ will project onto a
symplectic structure which provides an alternative symplectic
structure for $\wt{\Ga}$. In terms of the distributions $ker \om$ and
$ker \om_1$, the compatibility condition reads
$ker \om \subset ker \om_1$.
Again $ker \om = ker \om_1$ implies that $\om_1$ will project onto a
symplectic structure on the reduced manifold.

Of course one can start with $\om_1$ instead of $\om$ and the reduced
carrier space could be different. Then one looks for all other
symplectic structures that are going to be compatible with the given
reduction.

Summing up, we have found that if $\wt{\Ga}$ is the reduced dynamics
associated with $\Si$ and $\om$, any other symplectic description for
$\Ga$ will provide an alternative description for $\wt{\Ga}$ iff
$\F(\om_1,\Si) = \F(\om,\Si)$.

Let us consider an example to illustrate the situation.

\bigskip

\subsubsect{Reduction of free motion and symplectic
description}\label{se:fre}
On $T^*\Rb^3$ with standard symplectic structure $\om_0$
we consider free
motion provided by the Hamiltonian

\be
H={p^2\over 2m}={1\over 2m}\left ( p_{r}^2+{L^2\over r^2}\right )~,
\ee
where
\be
p_{r}={\vec p\cdot \vec r\over r}\quad,\quad \vec L=\vec p\wedge \vec r~.
\ee

We consider the invariant submanifold $\Si_{\vec c}$ obtained by
fixing the value of the angular momentum $\vec L$,
\be
\Sigma_{\vec c} =\left \{ (\vec r,\vec p)\vdash \vec L = \vec c
\right \}~.
\ee
The pull-back of $\om_0$ to $\Si_{\vec c}$ will be denoted by
$\om_{\vec c}$. In order to compute $ker\om_{\vec c}$ we consider the
infinitesimal generators $X_1, X_2, X_3$
of the rotation group on $T^*\Rb^3$. They are given by
\be
X_i = \varepsilon_{ijk}(x_j \pd{},{x_k} + p_j \pd{},{p_k})~,~~ i =
1, 2, 3~.
\ee
Now $ker\om_{\vec c}$ is generated by the Hamiltonian vector field
$Y$ associated with $\frac{1}{2}L^2$ when restricted to $\Si_{\vec c}$.
It is given by
\be
Y_{\vec c} = c_1 X_1 + c_2 X_2 + c_3 X_3~.
\ee
The reduced manifold $\wt{\Si}_{\vec c}$ is
defined as the submanifold of $\Rb^3$ endowed with coordinates
$\xi_1 = r^2, ~\xi_2 = p^2~, ~\xi_3 = \vec r \cdot \vec p$ ~, with the
relation ~$\xi_1 \xi_2 - \xi_3^2 = c^2$. It is diffeomorphic with
$T^*\Rb^+$ if we set $x = r~, ~p_x = (\vec p\cdot \vec r) / r$. The
reduced dynamics is Hamiltonian and we have
\bea
&&\wt{\Ga} = \frac{p_x}{m}\pd{},{x} +
\frac{c^2}{mx^3}\pd{},{p_x}~,\nn\\
&&\wt{H} = \frac{1}{2m}\left (p^2_x + {c^2\over x^2}\right )~,\nn\\
&&\wt{\om} = dp_{x}\wedge dx~.
\eea

There are many alternative symplectic descriptions for the free motion
on $T^*\Rb^3$. For instance,
\be
\om_F = d\left(\pd{F},{p_i} dq^i \right)~, ~~{\rm with}~
{}~det\vert\vert \pd{^2F},{p_i\partial p_j} \vert\vert \not= 0~,
\ee
provides a family of them. We shall consider a particular one just to
illustrate the procedure. We consider
\be
\om_1 = \om_0 + s d p^2 \wedge d(\vec p \cdot \vec r)~,
\ee
with $s$ a dimensional constant. The corresponding Hamiltonian for
the free motion is given by
\be
H_1 = \frac{p^2}{2m} + \frac{1}{2} s \frac{p^4}{m}~.
\ee
By restricting
$\om_1$ to $\Si_{\vec c}$ we get an alternative symplectic structure
for $\wt{\Ga}$ given by
\bea
\wt{\om}_1 &=& dp_{x}\wedge dx + s d (\frac{c^2}{x^2} + p_x^2) \wedge
d(x p_x) \nn \\
&=& \left(1 + 2s ( p^2_x + \frac{c^2}{x^2} ) \right) d p_x \wedge d x ~.
\eea
The corresponding Hamiltonian function is
\be
\wt{H}_1 = \left( 1 + s( p^2_x + {c^2\over x^2} )\right )
\frac{1}{2m}\left (p^2_x + {c^2\over x^2}\right )~.
\ee

Of course, any other admissible symplectic structure $\om'$ which
does not satisfy
\be
dL^2 \wedge (i_{L^j X_j} \om') = 0 ,
\ee
will not be projectable.
To satisfy the requirement that $\om'$ restricted to $\Si_{\vec c}$
is projectable onto $\wt{\Si}$, we need that the foliating
distribution is in the kernel of $\om' \vert_\Si$.
It may happen that the dimension of the kernel is too small for this
to happen. Therefore, to get
alternative symplectic structures for the reduced dynamics $\wt{\Ga}$
on $\wt{\Si}$ we shall also allow to start with degenerate two forms
on the initial carrier space as long as they are invariant under the
dynamical evolution.

Let us illustrate the situation.
For our free motion we consider the invariant submanifold defined by
\be
\Sigma_{\ell} = \left \{ (\vec r,\vec p)\vdash p^2 r^2 - r^2 p_r^2 =
\ell^2 \right \}~,
\ee
with equivalence relation provided by the rotation group. It is clear
that the kernel of a symplectic structure on a codimension one
submanifold must be one dimensional. It follows that there is no
symplectic structure that we can restrict to $\Si_\ell$ and that
project to $\wt{\Si}_\ell$. We have to start with a degenerate one.
For instance we could take
\be
\om = f_1(p^2, \vec p \cdot \vec r) dp^2 \wedge d(\vec p
\cdot \vec r) + f_2(p^2, r^2) dp^2 \wedge d r^2 +
f_3(r^2, \vec p \cdot \vec r) dr^2 \wedge d(\vec p \cdot \vec r)~.
\ee
In this case, to meet the invariance requirement under the
dynamical evolution, we would require that $L_{\Ga}\om = 0$ on
the submanifold $\Si_\ell$.
The choice $\om = -\frac{1}{2r^2} dr^2 \wedge d(\vec p \cdot \vec r)$
has a restriction to $\Si_\ell$ that is projectable onto
\be
\wt{\om} = dp_{r}\wedge  dr~.
\ee
with
\be
\wt{H} = {1\over 2m}\left ( p_{r}^2 + {\ell^2\over r^2}\right )~.
\ee
and
\be
\wt{\Ga} = {p_{r}\over m}{\partial\over \partial r} +
{\ell^2 \over mr^3} {\partial\over \partial p_{r}}~.
\ee

A similar situation occurs if we fix the energy, so getting the
invariant submanifold
$\Sigma_E = \{ (\vec p, \vec r) \vdash p^2 = 2mE\}.$
For the equivalence
relation we still use the rotation group. In this case the
reduced dynamics is given by
\be
\wt{\Ga} ={p_{r}\over m}{\partial\over \partial r}+\left ({2E\over r}-
{p_{r}^2\over mr}\right ){\partial\over \partial p_{r}}~,
\ee
which is Hamiltonian on $T^*\Rb^+$ with structures
\be
\wt{\om} = r^2 dp_{r}\wedge dr~,~~~~~
\wt{H} = Er^2- {r^2 p_{r}^2\over 2m}~.
\ee
\bigskip

\bigskip

Besides free motions on vector spaces, there are others dynamical systems
which can be easily integrated. They are geodesical motions on
Lie groups or on homogeneous spaces. Their flows are provided by the
action of one-parameter subgroups. Therefore, in the following
sections we are going to consider geodesical motions on these spaces.

\bigskip

\bigskip

\sect{Reducing geodesical motion on Lie groups}

\subsect{Few elements of Lagrangian formalism}
We briefly recall few elements of the geometry of the tangent bundle
[MFLMR].
Let $TQ$ be the tangent bundle to an $n$-dimensional
configuration space
$Q$, with local coordinates $\{q^i, u^i~,~i \in \{1,\dots\,n\}\}$.
On $TQ$ there are two natural tensor fields which essentially
characterize its structure. They are the
{\it vertical endomorphism} $S$
and the {\it dilation vector field} $\Delta$
which, in local coordinates are respectively given by
$S = dq^i\otimes {\partial \over {\partial u^i}}~$,
$\Delta = u^i \pd{~},{u^i} $~.
A {\it second-order derivation}, or SODE for short, is any vector
field $\Ga$ on $TQ$ such that
$S( \Gamma) = \Delta$.
Locally a SODE is of the form
$\Ga =~ u^i \pd{~},{q^i} + \Ga^i (q, u) \pd{~},{u^i}$.

There are two natural
lifting procedures for vector fields from $Q$ to $TQ$,
namely the tangent and the vertical lifting. If $X = X^i(q)
\pd{},{q^i} \in \X(Q) $, its {\it tangent lift} $X^T$ and its
{\it vertical lift} $X^V$ are the elements in $\vectq$ given by
\be
X^T = X^j(q) \pd{},{q^j} + u^i \pd{X^j},{u^i}\pd{},{u^j}~,~~~
X^V = X^j(q) \pd{},{u^j}~. \label{*grab1}
\ee

Let us take an element $\L \in {\F(TQ)}$. The $1$-form
$\theta_{\L} $ on $TQ$ defined by
$\theta_{\L} = d\L  \circ S~$
is the {\it Cartan 1-form} of the
{\it Lagrangian} $\L$. The Lagrangian $\L$ is said to be
regular, if the {\it Cartan $2$-form}
$\omega_{\L}$ on $TQ$ defined by
$\omega_{\L} = - d \theta_{\L}$~, is non degenerate.

Given a Lagrangian $\L$, the
{\it Euler-Lagrangian equations} for $\L$ are the following equations
\be
L_{\Gamma} \theta_{\L} - d \L = 0~,
\lb{*grab4}
\ee
where the unknown quantity is the vector field $\Gamma$.
In local coordinates \rf{*grab4} reads
\be
\tev q^i = u^i~,~~~~~
\fraz{d},{dt}\pd{\L},{u^i} - \pd{\L},{q^i} = 0~. \lb{*grab5}
\ee

If $\L$ is a regular Lagrangian, the solution of \rf{*grab4}
turns out to be a SODE. In this case
eqs.  \rf{*grab4}
can be written in an equivalent symplectic form by using the
{\it energy} $E_{\L}$ of $\L$,
\be
E_{\L} =: L_{\Delta} \L - \L = i_{\Gamma} \theta_{\L} - \L ~.
\lb{*grab6}
\ee
Then, by using Cartan identity, eqs.\rf{*grab4} are written as
\be
i_{\Gamma} \omega_{\L} = d E_{\L}~.
\lb{*grab7}
\ee

Sometimes it turns out to be useful to `project' equations \rf{*grab4}
along a non holonomic basis $X_1, \dots, X_n$ of vector fields on
$Q$. To do that one takes the contraction of \rf{*grab4} with the
corresponding tangent lifts $X^T_i \in \X(TQ)$. When
$X \in \X(Q)$, the contraction
$i_{X^T}(L_{\Gamma} \theta_{\L} - d \L) = 0 $
can be written as
\be
L_{\Gamma} i_{X^T}\theta_{\L} - L_{X^T} \L = 0~.
\lb{*grab9}
\ee
If we choose a basis of vector fields, we obtain a global expression
for the familiar Euler-Lagrangian equations.

\bigskip

\subsect{Geodesical motion on Lie groups}
Let $G$ be a Lie group thought of as a subgroup of the group of
matrices $GL(n,\Rb)$, and $\fkg$ its Lie algebra with
basis $\ta_1,\cdots,\ta_n~,~ [\ta_i, \ta_j] = c_{ij}^k \ta_k$.

On $G$ there is a canonical $\fkg$-valued left invariant $1$-form
\be
\alpha = \mcl~, \lb{ggro1}
\ee
which allows to define a basis $\{ \th^k \}$ of left invariant one forms in
$\X^*G$ via
\be
\mcl =: \ta_k \th^k~. \lb{ggro2}
\ee
These forms satisfy the Maurer-Cartan equation
\be
d\th^k + \fraz1,2 c_{ij}^k \th^i \wedge \th^j = 0~. \lb{ggro3}
\ee
The dual basis of left invariant vector fields $X_k$, defined by
\be
i_{X_k}\mcl = \ta_k~, \lb{ggro4}
\ee
satisfies $i_{X_k} \th^j = \de^j_k$, with commutations relations
\be
[X_i, X_j] = c_{ij}^k X_k~ \lb{ggro5}
\ee
along with
\be
L_{X_i}\th^k = - c_{ij}^k \th^j~. \lb{ggro5aa}
\ee

Analogously, one can construct right invariant forms $\wth^k$ and
vector fields $Y_k$, starting from the right invariant
$\fkg$-valued $1$-form $\beta = \mcr$,
\bea
&& \mcr =: \ta_k \wth^k~, \lb{ggro5a} \\
&& i_{Y_k} \mcr = \ta_k~. \lb{ggro5b}
\eea
Left and right invariant quantities are related by the adjoint
rapresentation whose matrix elements $D_i^j$ are defined by
\be
g^{-1}\ta_i g = D_i^j(g)~ \ta_j~. \lb{ggro5c}
\ee
In particular,
\be
\th^i = D^i_j(g) \wth^j~.
\ee

The left invariant vector fields are the generators of the right
action of $G$ on itself while the right invariant ones are the
generators of the left action. Therefore, they mutually commute
\be
[X_i, Y_j] = 0~. \lb{ggro6}
\ee

We can construct a basis of vector fields and $1$-forms
for the tangent group $TG$
starting with basis for $G$.
This is done as follows\footnote{With some abuse
of notation we shall identify $\ta_G^*\th^k$ with $\th^k$ and write
$i_{\Ga}\ta_G^*\th^k = \td{\th}^k$. }
\bea
&& \{X_k\} \mapsto \{(X_k)^v , (X_k)^T \}~, \nn \\
&& \{\th^k\} \mapsto \{\ta_G^*\th^k , d(i_{\Ga}\ta_G^*\th^k)\}~, \lb{ggro7}
\eea
with $\Ga$ any SODE on $TG$. Notice the the basis in \rf{ggro7} are
not dual to each other.

\bigskip

Let as suppose now that we have a metric $\me$ on $G$ which,
in the basis
of left and right invariant $1$-forms is written as
\bea
\me & = & \me_{jk} \th^{j} \otimes \th^{k}~, \lb{ggro8} \\
  & = & \wt{\me}_{jk} \wth^{j} \otimes \wth^{k}~,~~~~
\wt{\me}_{jk} = D_j^s D_k^t~ \me_{st}~. \lb{ggro8a}
\eea
The associated geodesical motion on $G$
is described by the Lagrangian
\be
\L_{\me} = \frac{1}{2}\tau_G^{*} \me ( \Ga, \Ga) =
\frac{1}{2} \me_{jk} \td{\th^{j}} \td{\th^{k}} ~, \lb{ggro9}
\ee
whose Cartan $1$-form turns out to be
\be
\thl = \me(\Ga,~ \cdot~ )~. \lb{ggro9a}
\ee
Right and left momenta associated with the respective actions
are given by
\bea
&& P^{(R)}_k =: i_{X^T_k} \thl = \me_{kj}\td{\th}^j~, \lb{ggro9b} \\
&& P^{(L)}_k =: i_{Y^T_k} \thl = \wt{\me}_{kj}\td{\wth^j}~. \lb{ggro9c}
\eea
The associated matrix valued momenta are given by
\bea
&& \P^{R} =: \me^{ij}\ta_i P^{(L)}_j = g^{-1}\td{g} ~, \lb{ggro9d} \\
&& \P^{L} =: \wt{\me}^{ij}\ta_i P^{(R)}_j = \td{g} g^{-1} ~. \lb{ggro9e}
\eea
Notice the $\P^{R}$ is invariant under {\it left} action while
$\P^{L}$ is invariant under the {\it right} one.

\bigskip

Since $TG$ is parallelizable in terms of left invariant (or right
invariant) vector fields, it turns out to
be more convenient to write the Euler-Lagrange equations
in the corresponding non holonomic
basis on $G$ as in \rf{*grab9}.
With the Lagrangian \rf{ggro9}, we have
$L_{Z^T} \L  =  \frac{1}{2} (L_{Z^T}\me)(\Ga, \Ga)$ so that
the Euler-Lagrange equations become
\be
\tev \me(\Ga, Z) = \frac{1}{2}(L_{Z^T}\me)(\Ga,\Ga)~.
\lb{ggro11}
\ee
As a consequence,
\be
\tev \me(\Ga, Z) = 0~~ \Longleftrightarrow ~L_{Z}\me = 0~.
\lb{ggro12}
\ee
The quantity $\me(\Ga,Z)$ can be thought of as the `angle' between
$\Ga$ and $Z$; it is a constant of the motion if and only if $Z$ is
a Killing vector field for $\me$.

In particular, the time evolution of the left and right momenta are
given by
\bea
&& \tev P_k^{(L)} = \frac{1}{2} (L_{Y_k^T}\me)(\Ga,\Ga)~,
\lb{ggro12a} \\
&& \tev P_k^{(R)} = \frac{1}{2} (L_{X_k^T}\me)(\Ga,\Ga)~,
\lb{ggro12b}
\eea
Therefore, left momenta are constants of the motion iff $\me$ is
left invariant and this is equivalent to the components $\me_{ij}$
being numerical constants.
As for the evolution of the right momenta, we find that
\be
L_{X_k^T}\me = -\me_{ij}c_{kl}^i(\th^j \otimes \th^l +
\th^l \otimes \th^j)~, \lb{ggro12ba}
\ee
and, after some algebra,
\be
\tev P_k^{(R)} = - c^i_{kr} \me^{rj} P_i^{(R)}
P_j^{(R)}~.\lb{ggro12bb}
\ee
Analogously, right momenta
are constants of the motion iff $\wt{\me}$ is
right invariant and this is equivalent to the components $\wt{\me}_{ij}$
being numerical constants. As for the evolution of
left momenta one would find an
expression similar to \rf{ggro12bb}.

We see that geodesical motion on
Lie group associated with left invariant (or right invariant) metrics
takes
the form of first order equations in the appropriate momenta.
In the language of momentum map, both left and right momenta give
maps
\be
\P^L~,~~\P^R~~ : T^*G \lra \fkg^*~, \lb{ggro12bc}
\ee
and the dynamics is projectable onto $\fkg^*$ where it is described
by a Hamiltonian vector field with respect to the
Konstant-Kirillov-Souriou Poisson structures (the $+$ and the $-$
structures) on $\fkg^*$ .

{}From the expressions \rf{ggro9d} and
\rf{ggro9e}, in each of the previous cases (left or right invariance)
the geodesical lines are just
translations of one-parameter subgroups of the group $G$,
\be
g(t) = g_0 e^{tA}~, ~~A \in \fkg~, \lb{ggro12c}
\ee
with $A$ the corresponding conserved velocity.

The converse is also true, namely any translation of a
$1$-parameter subgroup of $G$
represents a geodesical motion
with respect to a suitable metric on G. Let us suppose
we have a translation of a
$1$-parameter subgroup of $G$, $g(t) = B e^{tA}$. By writing
$\td{g} = A g$ we see that the left velocity $\mcdr = A$ is
a constant of the motion. Take now  any scalar product on the Lie
algebra $\fkg$, $n_{ij} =~ <\ta_i, \ta_j>$ and extend it to the unique
left invariant metric on $G$, given by
$\me = n_{ij} \th^i \otimes \th^j$.
The associated geodesical motion admits the starting
$1$-parameter subgroup as a possible motion. Had we written
$\td{g} = g A$ we should have extended the scalar product to the
unique right invariant metric. It should be noticed, however, that in
general it is not possible to have a bi-invariant metric on $G$
since it is not possible to have an adjoint invariant scalar product
on $\fkg$ in general. This is however true if $G$ is compact or semisimple.

\bigskip

Let us write the metric in its diagonal form.
Then we have
\bea
&& \me = \th^1 \otimes \th^1 + \cdots + \th^n \otimes \th^n~, \\
&& \L_{\me} =~ \fraz1,2
(\td{\th}^1 \td{\th}^1 + \cdots + \td{\th}^n \td{\th}^n)~, \\
&& \th_\L = \td{\th}^1 {\th}^1 + \cdots + \td{\th}^n {\th}^n~.
\lb{ggro13}
\eea

If the left invariant
vector fields $X_{k+1}, \cdots ,X_n$
are Killing vectors,
also their commutators are Killing vectors; therefore we can assume
that they close on a Lie subalgebra. We denote by $K$ the
corresponding subgroup in $G$. The associated momenta
$P^{(R)}_j~,~ j \in\{k+1, \cdots, n\}$
will be constants
of the motion for the geodesical motion.
If we fix their values, say $P^{(R)}_j = c_j$, we get a
submanifold $\Si_{\bf c}$ of $TG$.
We can consider now the pullback of $\om_\L$ to $\Si_{\bf c}$ to get
$\om_{\bf c}$. Its kernel $ker\om_{\bf c}$
determines an involutive distribution $\D$
on ${\Si_{\bf c}}$ and we get a reduced space
$\wt{\Si}_{\bf c} = \Si_{\bf c} / \D$
diffeomorphic to $T(G/K)$. As for the reduced
dynamics we can make few general considerations. The energy
\be
E_{\bf c} = \frac{1}{2} \sum_{i=1}^k (\td{\th^i})^2 + \frac{1}{2}
\sum_{k+1}^n (c_j)^2~, \lb{ggro15}
\ee
will be projectable on $T(G/K)$. We can associate a metric tensor to
it by setting $\me_{\bf c} = \sum_{i=1}^k \th^i \otimes \th^i$.
We consider also the metric Lagrangian
$\L_{\bf c} = \frac{1}{2} \sum_{i=1}^k (\td{\th^i})^2$ and the associated
$2$-form $\om_{\L_{\bf c}}$. We see that if we consider the pullback
of $\om_{\L_{\bf c}}$ to $\Si_{\bf c}$ and compare it with
$\om_{\bf c}$ we find that they differ
by a term containing $2$-forms like $c_j d \th^j$. Since our dynamics
on $\Si_{\bf c}$ satisfies both equations
$i_{\Ga_{\bf c}} \om_{\L_{\bf c}} = dE_{\bf c}$ and
$i_{\Ga_{\bf c}} \om_{\bf c} = dE_{\bf c}$~, this implies that
the additional forces,
with respect to the geodesical ones on $G/K$, represent gyroscopic
forces or forces of the magnetic type. In the latter case, the values
of the $c_j$'s may be interpreted as contributing a `magnetic
charge'.

\bigskip

We are going to illustrate these general considerations on a familiar
example in the following section.

\bigskip

\subsect{Geodesical motion on $SU(2)$}\label{se:geo}
As an example we consider geodesical motion on $SU(2)$ and project it
onto motions on the $2$-dimensional sphere.

We think of $SU(2)$ in terms of unitary $2\times 2$ matrices with
determinant equal to one. As a basis for its Lie algrebra we take
$\ta_k = \frac{i}{2}\si_k,~, k \in \{1,2,3\} $ the three Pauli
matrices. Then $[\ta_j, \ta_k] = \varepsilon_{jk}^l \ta_l$.

Given any triple $(n_1, n_2, n_3) \in \Rb^3$, we consider
the left-invariant metric on $SU(2)$ given by
\be
\me = n_1 \th^1 \otimes \th^1 + n_2 \th^2 \otimes \th^2
+ n_3 \th^3 \otimes \th^3~. \lb{ggro15a}
\ee
The associated Lagrangian
\be
\L = \fraz1,2
(n_1 \td{\th}^1 \td{\th}^1 + n_2\td{\th}^2 \td{\th}^2 +
n_3\td{\th}^3 \td{\th}^3)~ \lb{ggro15b}
\ee
has the following  Euler-Lagrange equations:
\bea
&&\tev P_k^{(L)} = 0~,~~~k \in \{1, 2, 3 \}~, \nn \\
&&\tev P_1^{(R)} = \frac{n_2 - n_3}{n_2n_3} P_2^{(R)} P_3^{(R)}~, \nn \\
&&\tev P_2^{(R)} = \frac{n_3 - n_1}{n_3n_1} P_3^{(R)} P_1^{(R)}~, \nn \\
&&\tev P_3^{(R)} = \frac{n_1 - n_2}{n_1n_2} P_1^{(R)} P_2^{(R)}~,
\eea
namely the Euler equations for a top.

\bigskip

On $SU(2)$  there is a natural bi-invariant metric corresponding to
the Cartan-Killing form on its Lie algebra which, when dealing with
matrices, can be expressed in terms of the trace.
The associated Lagrangian is given by
\be
\L = tr \td{g}(\td{g^{-1}}) = - tr \mcdl \mcdl = - tr \mcdr \mcdr~,
\lb{ggro16}
\ee
which in terms of left or right invariant forms is
\be
\L = \fraz1,2
(\td{\th}^1 \td{\th}^1 + \td{\th}^2 \td{\th}^2 +
\td{\th}^3 \td{\th}^3) =
\fraz1,2
(\td{\wth^1} \td{\wth^1} + \td{\wth^2} \td{\wth^2} +
\td{\wth^3} \td{\wth^3})~. \lb{ggro17}
\ee
The associated geodesical equations of the motion are
\be
\tev \mcdl = 0~,~~~~~\tev \mcdr = 0~, \lb{ggro17a}
\ee
or, equivalently,
\be
\tev \td{\th}^k = 0~,~~~~~\tev \td{\wth^k} = 0~,~~~k \in \{1,2,3\}~.
\lb{ggro17b}
\ee

Let us consider now the projection
\be
\pi : SU(2) \ra S^2~, ~~~ g \mapsto
g \sigma_3 g^{-1} =: \vec{x} \cdot \vec{\si}~. \lb{ggro18}
\ee
One can verify that $\vec{x} \cdot \vec{x} =1$ so that $x^1, x^2, x^3$ are
coordinates on $S^2$. By identifying them with their pullback to
$SU(2)$, the Lagrangian \rf{ggro17} and the associated Cartan
$1$-form $\thl$ can be written as
\bea
&&\L = \frac{1}{2} \left(
(\td{x}^1)^2 + (\td{x}^2)^2 + (\td{x}^3)^2 \right) +
\frac{1}{2} (\td{\th}^3)^2~, \lb{ggro18a} \\
&& \thl = \td{x}^1 d x^1 + \td{x}^2 d x^2 + \td{x}^3 d x^3 +
\td{\th}^3 \th^3 ~. \lb{ggro18b}
\eea

Let us now inquire about the projectability of the geodesical motion
\rf{ggro17a}-\rf{ggro17b} to $TS^2$. We consider
\bea
&& \td{x}^j = \tev x^j~, \\
&& \ttd{x}^j = \tev \td{x}^j~, \lb{ggro19}
\eea
where $\tev$ is the time-derivative along the geodesical motion on
$SU(2)$. This
motion would be projectable if $\ttd{x}^j$ could be expressed in terms
of the variables $\td{x}^j$ and $x^j$ alone.
In terms of left invariant forms, after some algebra one arrives at
\bea
&& \td{\vec{x}} \cdot \vec{\si} =
g (\td{\th}^2\si_1 - \td{\th}^1\si_2) g^{-1}~, \lb{ggro20} \\
&& \ttd{\vec{x}} \cdot \vec{\si}
= - ((\td{\th}^1)^2 + (\td{\th}^2)^2) g \si_3 g^{-1} +
\td{\th}^3 g (\td{\th}^1\si_1 + \td{\th}^2\si_2) g^{-1}~. \lb{ggro21}
\eea

We find that,
in general, the motion \rf{ggro21} is not projectable onto a
motion on $S^2$. The
term that cannot be expressed in terms of quantities on $TS^2$ is
$\td{\th}^3$. However, since $\td{\th}^3$ is a constant of the
motion, we can fix a value for it which determines an invariant
submanifold $\Si$
of $TSU(2)$ which covers $TS^2$. The motion restricted to $\Si$ is
projectable to $TS^2$.
For each submanifold $\Si$ we find a different dynamical system on
$TS^2$ which is parametrized by the values of $\td{\th}^3$.
We find two classes of motions which are quite
different depending on whether $\td{\th}^3$ vanishes or not:

\bigskip

\noindent
1.~~ $\td{\th}^3 = 0$.

\noindent
The resulting motion is the geodesical motion on $S^2$. From
\rf{ggro20} we get that
\be
(\td{\th}^1)^2 + (\td{\th}^2)^2 = (\td{x}^1)^2 + (\td{x}^2)^2 +
(\td{x}^3)^2~. \lb{ggro22}
\ee
As a consequence,
\be
\tev \td{\vec{x}} = - (\td{\vec{x}}^2) \vec{x} ~. \lb{ggro23}
\ee

\bigskip

\noindent
2.~~ $\td{\th}^3 = k \not= 0$.

\noindent
The resulting motion is the motion of a charged particle on $S^2$
moving in the field of a magnetic monopole situates at the center of
the sphere. To better see this fact let us start again from the
Lagrangian \rf{ggro18a}. We find that
\bea
&& \thl \vert_{\Si} = \td{x}^1 d x^1 + \td{x}^2 d x^2 + \td{x}^3 d x^3 +
k \th^3 \vert_{\Si}~, \lb{ggro24} \\
&& \oml \vert_{\Si} = d x^1 \wedge d \td{x}^1  +
d x^2 \wedge d \td{x}^2 + d x^3 \wedge d \td{x}^3 -
k d \th^3 \vert_{\Si}~. \lb{ggro25}
\eea
Now, $\th^3$ is just the monopole connection on the $U(1)$ bundle
$SU(2) \ra S^2$ with $d \th^3$ the corresponding curvature. It turns
out that the latter is the volume form on $S^2$,
\be
d\th^3 = \frac{1}{2}\varepsilon_{jlm} x^j d x^l \wedge d x^m~. \lb{ggro26}
\ee
The $2$-form \rf{ggro25} becomes
\be
\oml \vert_{\Si} = d x^1 \wedge d \td{x}^1  +
d x^2 \wedge d \td{x}^2 + d x^3 \wedge d \td{x}^3
- \frac{k}{2}\varepsilon_{jlm}~ x^j d x^l \wedge d x^m~.
\lb{ggro27}
\ee
As for the energy, we find
\be
E_{\L} \vert_{\Si} = \frac{1}{2} \left(
(\td{x}^1)^2 + (\td{x}^2)^2 + (\td{x}^3)^2 \right) +
k^2~. \lb{ggro28}
\ee
The dynamical system determined by the couple
$(\oml \vert_{\Si}, E_{\L} \vert_{\Si})$ is found to be
\be
\wt{\Ga} = \td{x}^j \pd{},{x^j} - k \varepsilon_{jlm}~ x^j \td{x}^l
\pd{},{\td{x}^m}~, \lb{ggro29}
\ee
with equations of the motion given by
\be
\tev \td{x}^l = - k \varepsilon_{ljm}~ x^j \td{x}^m~. \lb{ggro30}
\ee

\bigskip

\bigskip

\sect{The Calogero-Moser system}
We may parametrize elements in $\Rb^3$ using $2\times2$ symmetric
matrices
\be
\Xb = \left\{ \begin{array}{cc}
x_1 & \fraz1,{\sqrt2} x_2 \\
\fraz1,{\sqrt2} x_2 & x_3
\end{array} \right\}~. \lb{reso1}
\ee

Free motion on $T\Rb^3$ can be written as
\be
\ttd{\Xb} = 0~, \lb{reso2}
\ee
with standard Lagrangian function
\be
\L = \fraz1,2 tr {\td \Xb}^2 = \fraz1,2 ({\td x}_1^2 + {\td x}_2^2 +
{\td x}_3^2)~. \lb{reso21}
\ee
The matrix
\be
\Mb = [\Xb, \td{\Xb}]~\lb{reso3}
\ee
is then a constant of the motion. Since $\Mb$ is an antisymmetric matrix
it can be written as
\be
\Mb = \ell~ {\bf \sigma}~,~~~~~ {\bf \sigma} = \left\{ \begin{array}{cc}
0 & 1 \\
-1 & 0
\end{array} \right\}~, \lb{reso4}
\ee
with $\ell$ the absolute value of the total angular momentum.

We shall now show how to reduce the dynamics \rf{reso2} to the
Calogero-Moser dynamics on $T\Rb^2$. We recall that this dynamics is
associated with the following Lagrangian on $T\Rb^2$ (see for
instance [Mo])
\be
\L_{CM} = \fraz1,2 (\td{q}_1^2 + \td{q}_2^2)
- \fraz{\ell^2},{(q_2 - q_1)^2}~. \lb{reso4a}
\ee

\bigskip

Since $\Xb$ is a symmetric matrix,
it can be diagonalized by elements in the
rotation group $SO(2)$ by a similarity transformation,
\be
\Xb = \Gb~ \Qb~ \Gb^{-1}~, \lb{reso5}
\ee
with
\be
\Qb = \left\{ \begin{array}{cc}
q_1 & 0 \\
0 & q_2
\end{array} \right\}~,~~~~~
\Gb = \left\{ \begin{array}{cc}
cos\varphi & sin\varphi \\
-sin\varphi & cos\varphi
\end{array} \right\}~. \lb{reso6}
\ee
As a consequence,
\be
\td{\Xb} = [\td{\Gb}~ \Gb^{-1}, \Xb] + \Gb~ \td{\Qb}~ \Gb^{-1} =
\Gb \{[\Gb^{-1}~ \td{\Gb}, \Qb] + \td{\Qb}~ \} \Gb^{-1}~
\lb{reso7}
\ee
and
\be
\Mb = [\Xb, [\td{\Gb}~ \Gb^{-1}, \Xb]] =
\Gb~ [\Qb, [\Gb^{-1}~\td{\Gb}, \Qb]] \Gb^{-1}~. \lb{reso8}
\ee
Using the fact that
\be
\Gb^{-1}~ \td{\Gb}~ =~ \td{\Gb}~ \Gb^{-1}~
=~ \td{\varphi} \sigma~, \lb{reso9}
\ee
after some algebra, we can derive the $\ell$ in \rf{reso4} as
\be
\ell = - \fraz1,2 tr \Mb \sigma = \td{\varphi}(q_2 - q_1)^2~.
\lb{reso10}
\ee

By deriving \rf{reso7} once more with respect to time, after some
algebra we get the following equation
\be
\ttd{\Qb} - \td{\varphi}^2 [\sigma, [\sigma, \Qb]] = 0~. \lb{reso101}
\ee
{}From which,
if we derive $\td{\varphi}$ from \rf{reso10}, we get the equations
\bea
&& \ttd{q}_1~ = - \fraz{2\ell^2},{(q_2 - q_1)^3}~, \nn \\
&& \ttd{q}_2~ = \fraz{2\ell^2},{(q_2 - q_1)^3}~. \lb{reso11}
\eea

By using the fact that
$\ell$ is a constant of the motion,
for each value $\ell = const$ we find an invariant submanifold
$\Si_{\ell}$ in the tangent bundle of the space of symmetric matrices,
which cover $T\Rb^2$. Projecting from each one of these submanifolds
we find a family of dynamical systems for
the variables $q_1, q_2$
which we can read as a second order dynamical systems on $T\Rb^2$.
We have two classes:

\noindent
1.~~~$\ell = 0$~.~~~Free motion on $\Rb^2$.

\noindent
2.~~~$\ell \not= 0$~.~~~Calogero-Moser systems on $\Rb^2$.

\bigskip

\noindent
{\bf Remark.}
Equations \rf{reso11} on $T\Rb^2$ admit a
Lagrangian description with Lagrangian function given by \rf{reso4a}.
However, the latter cannot
be gotten by reduction of the starting free Lagrangian \rf{reso21}.
Indeed, using the coordinates $q_1, q_2$ and $\varphi$, the
Lagrangian \rf{reso21} can be written as
\bea
\L &=& \fraz1,2 tr\td{\Gb}^2 + \fraz1,2 tr [\Gb^{-1}\td{\Gb}, \Qb]^2
\nn \\
&=& \fraz1,2 (\td{q}_1^2 + \td{q}_2^2) + \fraz{\ell^2},{(q_2 -
q_1)^2}~.
\eea
If we fix now the value of $\ell$ to be a constant one, the resulting
function is a Lagrangian on $T\Rb^2$ which does not coincide with
\rf{reso4a} since the
potential term has the opposite sign.

\bigskip

\subsect{~Symplectic ~reduction ~and ~deformation ~for ~the ~
Calogero-Moser system}
Let us start with the canonical symplectic structure $\omega$ on
$T^*\Rb^3$ given in canonical coordinates $(x,y)$ by $\omega =
dy_i\wedge dx_i.$
Identifying $\Rb^3$ with the space of symmetric matrices as in
\rf{reso1},
consider the symplectic action of the group $SO(2)$ on this cotangent
bundle implemented by the action on $\Rb^3$ defined by
${\bf X}\mapsto {\bf G}{\bf X}{\bf G}^{-1},$
for $\bf G$ being as in
\rf{reso5}.
The infinitesimal generator $Y$ of this action is a Hamiltonian
vector field with the Hamiltonian
\be
p_0=\sqrt2(x_2(y_1-y_3)+(x_3-x_1)y_2)~. \lb{mose1}
\ee
Introducing new coordinates $(\varphi ,q_1,q_2)$ in $\Rb^3$
(cf. \rf{reso6}) by
\bea
&&x_1=q_1cos^2\varphi +q_2sin^2\varphi~, \nn \\
&&x_3=q_1sin^2\varphi +q_2cos^2\varphi~, \nn \\
&&x_2={sin2\varphi \over \sqrt2}(q_2-q_1)~, \lb{mose2}
\eea
we get conjugate coordinates in the cotangent bundle of the form
\bea
&&p_\varphi =p_0=(q_2-q_1)(y_1-y_3)sin2\varphi +\sqrt2
(q_2-q_1)y_2cos2\varphi~, \nn \\
&&p_{q_1}=p_1=y_1cos^2\varphi +y_3sin^2\varphi
-y_2{sin2\varphi \over \sqrt2}~, \nn \\
&&p_{q_2}=p_2=y_1sin^2\varphi +y_3cos^2\varphi
+y_2{sin2\varphi \over \sqrt2}~, \lb{mose3}
\eea
i.e. $\omega =dp_i\wedge dq_i,$ were we put $q_0=\varphi .$
In the new coordinates $Y=\partial _\varphi .$
The standard reduction with respect to the action of $SO(2)$ is then
the following.
For a regular value $k$ of the momentum mapping $p_0$ consider the
submanifold $\Sigma _k=p_0^{-1}(k).$ Then the subbundle of kernels
of $\omega _{\mid {\Sigma _k}}$  in $T\Sigma _k$ is generated by $Y$
and we can pass to the quotient manifold $(\wt{\Sigma} _k,
\wt{\omega} ).$
Since $(q_1, q_2, p_1, p_2)$ are clearly generators of the algebra of
$Y$--invariant functions on each $\wt{\Sigma}_k$,
we can think of $\wt{\Sigma} _k$ as $T^*\Rb^2$ with
canonical coordinates $(q,p).$ We can also reduce to
$\wt{\Sigma} _k$ any dynamics represented by a $Y$--invariant
Hamiltonian on $T^*\Rb^3.$ For example, the Hamiltonian
$H_0={1\over 2} \sum_i y_i^2$ of the free motion $\Gamma$
is $Y$--invariant and we can
write it in the new coordinates in the form
\be
H_0={1\over 2}(p_1^2+p_2^2)+{p_0^2\over 4(q_2-q_1)^2}~, \lb{mose4}
\ee
so the reduced Hamiltonian describes the Calogero--Moser system
\rf{reso11}
with $\ell={p_0\over 2}.$
We can also reduce all $Y$--invariant potentials, i.e. potentials
of the form
\be
V=V(x_1+x_3,(x_1-x_3)^2+2x_2^2)~. \lb{mose5}
\ee
The reduced potential
is then $\wt{V} (q_1,q_2)=V(q_1+q_2,(q_2-q_1)^2).$
\smallskip
Observe now that we can reduce to $\wt{\Sigma} _k$ not only $\omega$
but any symplectic form $\omega '$ such that
$\omega '=\alpha \wedge dp_0+\omega '',$ where $\alpha$ is any
1--form and $i_Y\omega ''=0.$
Consider now functions
\bea
&& P=p_1+p_2=y_1+y_2~, \nn \\
&& F=(p_1-p_2)^2+{p_0^2\over (q_2-q_1)^2}=(y_3-y_1)^2+2y_2^2~,
\eea
and the 2--form
\be
\eta =dp_2\wedge dq_1+dp_1\wedge dq_2~.
\ee
They are clearly
projectable and one can check that
\bea
&& i_\Gamma \eta =dG~, \nn \\
&& G=p_1p_2-{p_0^2\over 4(q_2-q_1)^2}~.
\eea
Then, if we put now
\be
\omega_{\lambda,s,t}= \lambda \omega +s\eta +tdP\wedge dF~,
\ee
we get a closed projectable and non--degenerate (at least for
$\lambda \not=0 $ and small $s$) 2--form,
i.e. a new projectable symplectic structure.
Our free motion $\Gamma$ can be now equivalently described by the
Hamiltonian $H_{\lambda,s}= \lambda H_0+sG$
with respect to the symplectic form
$\omega _{\lambda,s,t}$ ($F, P$ are constants of the motion).
All this project to $\wt{\Sigma} _k,$
so the Calogero--Moser dynamics $\wt{\Ga}$ can be equivalently described by
a three--parameter family of symplectic structures
$\omega _{\lambda,s,t}$ on $T^*\Rb^2$
and two--parameter family of corresponding Hamiltonians $H_{\lambda,s}$.
\smallskip
Observe that  the form $\omega _{0,s,t}$ is degenerate but its
projection $\wt{\omega}_{0,s,t}$ is symplectic for $s \not= 0$, so
that $s\wt{G}$ is also an alternative Hamiltonian for $\wt{\Ga}$ and
$\omega _{0,s,t}$.

The Poisson structure $\wt{\Lambda} _{\lambda,s,t}$ on $T^*\Rb^2$
corresponding to $\wt{\omega}_{\lambda,s,t}$
is given by
\bea
\wt{\Lambda} _{\lambda,s,t} &=&
\frac{1}{\lambda^2 - s^2} \left(
(\lambda - tf ) \partial_{p_1}\wedge \partial _{q_1} -
(s - tf ) \partial_{p_2}\wedge \partial _{q_1} \right. \nn \\
&& ~~~~~~~~~~~~~~~~~~ \left. -(s + tf) \partial_{p_2}\wedge \partial _{q_1}
+(\lambda + tf) \partial_{p_2}\wedge \partial _{q_2} \right)~,
\lb{mose6}
\eea
where $f = \frac{2p_0^2}{(q_2 - q_1)^3}$ .

\bigskip

By using the generalization of the Calogero-Moser system to
$n$-particles described in [OP], we can extend our considerations to
$n$-particles. In particular we can exhibit alternative Lagrangian,
Hamiltonian and Poisson structures like in the case we have
considered.

\bigskip

\bigskip

In the previous examples we have studied a reduced motion along
homogeneous spaces of the form $M / K$. In the following examples we
shall consider a complementary situation in the splitting $K \ra M
\ra M /K$. We shall consider the projection of the motion on $M$
along the leaves of the foliation. In this way $M/K$ will parametrize
a family of dynamical systems.

\bigskip

\bigskip

\sect{Reduction of geodesical motions}
We shall first consider the simple case of geodesical motion on the
Euclidean spheres and then we shall treat the more general case of
geodesical motion on general hyperboloids.

\bigskip

\subsect{Geodesical motions on spheres}
The geodesical motion on $S^2$ can also be obtained by reducing a
dynamical system different by the one considered in
section \ref{se:geo}. The setting can be easily generalized to
the $n$-dimentional sphere so we shall describe the general case.

On $T^*\Rb^{n+1}$ with standard symplectic structure
$\omega =dp_i\wedge dq^i$, we consider the following Hamiltonian
\be
H = {1\over 2}Gram(\vec p, \vec q) =
\frac{1}{2}\left( {p}^2 {q}^2 - (\vec p \cdot \vec q )^2
\right)~. \lb{j1}
\ee
The corresponding equations of motion can be integrated by
exponentiation. Indeed, on any submanifold
that we obtain by fixing the values of
${p}^2, {q}^2$ and $ \vec p \cdot \vec q $ (they are
constants of the motion) the dynamics becomes linear
(depending on the specific values of the constants of the motion).
Our system is then a simple system.
We fix now the submanifold $\Sigma
=\{(\vec p, \vec q)\vdash {q}^2 =1\}$ and
as an equivalence
relation on $\Sigma $ we take $({\vec p}_1, {\vec q}_1)
\R ({\vec p}_2, {\vec q}_2)$
if ${\vec q}_1 = {\vec q}_2 $ and ${\vec p}_1
= {\vec p}_2 + s {\vec q}$, for some $s\in \Rb$.
It is possible to embedd the quotient manifold into $T^*\Rb^{n+1}$
by selecting in every equivalence class an element $\vec{\pi}$
such that $\vec{\pi} \cdot \vec q =0$. This means
that in the equivalence class
of $(\vec p, \vec q)$ we consider $\vec{\pi} = \vec p -(\vec p\cdot
\vec q)\vec q$ with $q^2=1$. In this way the quotient manifold
becomes $T^*S^n$. The starting dynamical vector field
\be
\Gamma = \left(q^2\vec p - (\vec q\cdot \vec p) \vec q\right) \cdot
{\partial \over \partial \vec q}+
\left( (\vec q\cdot\vec p) \vec p - p^2\vec q\right) \cdot
{\partial\over \partial\vec p}~,
\ee
on the submanifold $T^*S^n$ becomes
\be
\wt{\Ga} =\vec{\pi} \cdot {\partial\over \partial \vec q } -
\pi^2 \vec q \cdot {\partial \over \partial \vec{\pi}}~.
\ee
The corresponding equations of motion in the second order form
\be
\ddot {\vec q}=-(\dot q )^2\vec q~,
\ee
represent the geodesical flow on the unitary sphere $S^n$.

\bsk

This example is to be kept in mind when addressing the unfolding
procedure: the same dynamical system can be unfolded in different
ways.

\bigskip

\subsect{Geodesical motions on hyperboloids}
In our example on the geodesical motion on $S^n$ we have started with
the Hamiltonian function \rf{j1}
on the cotangent bundle $T^*\Rb^{n+1}$. We have used the Euclidean
metric to lower and raise indices so as to consider $\vq$ and $\vp$ as
vectors having the same (co-) variance transformation properties.

Since we would like to go beyond the Euclidean metric, we cannot
assume $\vq$ and $\vp$ to be vectors with identical transformation
properties if they are going
to provide a symplectic chart for the symplectic structure we are
going to use. In order to avoid this problem we shall work on the
tangent bundle $T\Rb^{n+1}$ instead of the cotangent bundle.
However, we do not have any Lagrangian corresponding to the
Hamiltonian \rf{j1}  which we could use to construct a corresponding
Lagrangian formalism on $T\Rb^{n+1}$. What seems to be a natural
choice, namely
\be
\L = \frac{1}{2}\left( \td{q}^2 q^2 - (\vq \cdot \tvq)^2 \right)~,
\lb{j2}
\ee
turns out to be singular and it does not give rise to a symplectic
structure on $T\Rb^{n+1}$. We are forced therefore, to work on
$T\Rb^{n+1}$ with  a symplectic structure defined autonomously, even
though in a non natural way. We are going to use the so called symplectic
formalism for the tangent bundle [MSSV]. This means that there is no
connection between the generating function of the dynamics and the
Legendre map which is used to pull-back to the tangent bundle the
canonical symplectic structure on the cotangent bundle.

Let us start then with $T\Rb^{n+1} \equiv \Rb^{n+1} \times \Rb^{n+1}$
with coordinates $(\bq, \tbq) = (q^i, \td{q}^i)$ and consider any non
degenerate quadratic form $\me$ on $\Rb^{n+1}$,
\be
\me(u,v)=u^i v^j \me(e_i,e_j) = u^i v^j \me_{ij}~,~~~\me_{ij}
\in \Rb~,~~~det\vert\me_{ij}\vert \not=0~. \lb{j3}
\ee
We can construct a symplectic form $\om_{\me}$ and a function
$E_{\me}$ (generating function for the dynamics) by
\bea
&&\om_{\me} = \me(d\bq ~{\buildrel \wedge \over ,}~ d\tbq)
= \me_{ij}dq^i\wedge
d \td{q}^j~, \lb{j4} \\
&&E_{\me} = \frac{1}{2} \left( \me(\bq,\bq) \me(\tbq,\tbq) -
\me(\bq,\tbq)^2 \right)~. \lb{j5}
\eea
The dynamical vector field $\Ga$ determined by $i_{\Ga}\om_{\me} =
dE_{\me}$ turns out to be
\be
\Ga_{\me} = \left ( \me(\bq,\bq) \td{q}^j - \me(\bq,\tbq) q^j \right)
\pd{},{q^j} +  \left ( \me(\bq,\tbq) \td{q}^j - \me(\tbq,\tbq) q^j \right)
\pd{},{\td{q}^j}~. \lb{j6}
\ee
The three quantities $\me(\bq,\bq)$, $ \me(\bq,\tbq) $ and
$\me(\tbq,\tbq)$ are constants of the motion for $\Ga$. On any
submanifold $\Si$ of $T\Rb^{n+1}$ determined by
$\me(\bq,\bq) = c_1$, $ \me(\bq,\tbq) = c_2 $,
$\me(\tbq,\tbq) = c_3$~, $c_1, c_2, c_3 \in \Rb$, the dynamics
becomes linear and can be integrated by exponentiation, i.e.
it is a simple system.

We identify a particular submanifold by setting $c_1 = 1, c_2 = 0$
and we get a restricted vector field
\be
\wt{\Ga}_{\me} = \td{q}^j \pd{},{q^j} - \me(\tbq,\tbq)~
q^j \pd{},{\td{q}^j}~, \lb{j7}
\ee
giving
\be
\ttd{q}^j = - \me(\tbq,\tbq)~q^j~, \lb{j8}
\ee
namely, the geodesical motion on the `pseudo-sphere' (hyperboloid) in
$\Rb^{n+1}$ determined by $\me(\bq,\bq) = 1$.

\bigskip

\noindent
{\bf Remark.} As $\me(\tbq,\tbq)$ is a constant of the motion, on each
level set determined by $\me(\tbq,\tbq)=c_3$, we can still reduce the
dynamics \rf{j7} to
\be
\wt{\wt{\Ga}}_{\me} = \td{q}^j \pd{},{q^j} - c_3 q^j \pd{},{\td{q}^j}~,
\ee
which represents $n$ isotropic
harmonic oscillators.

\bigskip

If we select $\me(\bq,\bq) > 0$ and the orbit goes through the point
$\sqrt{\me(\bq,\bq)} (1, 0, \cdots, 0)$, the reduced space can be
identified with
\be
T^*\left(SO(p,(n+1)-p) / SO(p-1,(n+1)-p)\right)~, \lb{j9}
\ee
where $(p,(n+1)-p)$ is the signature of $\me$. The reduced
motion takes place on a homogeneous space which carries an action of
$SO(p,(n+1)-p)$  .

\bigskip

\noindent
{\bf Remark.}
We would like to mention that one could start with $\me(X,Y) =
X^{\dagger} M Y$ where $X,Y \in \Cb^{n+1}$ and $M$ is an Hermitiam
matrix, $M^{\dagger} = M$. The superscript $^{\dagger}$ denotes
Hermitian conjugation. In this case we would consider the group
$SU(p,q)$ acting on $\Cb^{n+1}$ and proceed to define homogeneous
spaces like in the real case. Systems based on these spaces have been
considered in [ORW].

\bigskip

By looking at our construction of $\om_{\me}$, we notice
that the Lagrangian function
$\L_{\me} = \frac{1}{2}\me_{ij} \td{q}^i\td{q}^j$
gives rise exactly to our symplectic structure as $\om_{\me} =
-d(\pd{\L_{\me}},{\td{q}^j} dq^j)$. It is therefore natural to
consider the free motion associated with  $\L_{\me}$ and to compare it
with the reduced geodesical motion on the hyperboloids. We notice
that one has also $E_{\L} = \td{q}^j \pd{\L},{\td{q}^j} - \L_{\me} =
\frac{1}{2}\me_{ij} \td{q}^i\td{q}^j$~.
We can apply the same decomposition we used for the free particle
motion in $\Rb^3$ in section \ref{se:fre} and write
\be
E_{\L} = \frac{1}{2} \tbq^2 + \frac{1}{2}
\frac{(\bq \wedge \tbq)^2}{\bq^2}~. \lb{j10}
\ee

Since our quadratic form is not necessarily positive definite and
we are not in three dimensions,we have to say what we mean by $\tbq^2$
and what is the meaning of $(\bq \wedge \tbq)^2$. To make
things similar to the three dimensional case we have to digress a
little to introduce the appropriate calculus.

\bigskip

\noindent
{\it Digression: Calculus with vector valued forms.}

We start with our non degenerate quadratic form on $\Rb^{n+1}$,
\be
\me(u,v)=\me(e_i,e_j) e^i \otimes e^j ~,~~~ \me_{ij} =
\me(e_i,e_j)~. \lb{j11}
\ee

By using vector valued forms, we can write
\bea
&& \bq \wedge d\bq = q^i d q^j e_i \wedge e_j =
\frac{1}{2} (q^i d q^j - q^j d q^i) e_i \wedge e_j~, \\  \lb{j12}
&& \me(d\bq ~{\buildrel \otimes \over ,}~ d\bq)
= d q^i \otimes d q^j \me(e_i, e_j) =
d q^i \otimes d q^j \me_{ij} ~, \\ \lb{j13}
&& \me(\bq \wedge d\bq ~{\buildrel \otimes \over ,}~ \bq \wedge d\bq) =
\frac{1}{4} (q^i d q^j - q^j d q^i)
\otimes (q^m d q^n - q^n d q^m) \times \nn \\
&&~~~~~~~~~~~~~~~~~~~~~~~~~~~~~~~~~~~~~~~~~~~~~~~~
\times \me(e_i \wedge e_j, e_m \wedge e_n )~,
\lb{j14}
\eea
where $\me(v \wedge u, w \wedge z )$ is the natural extension of the
quadratic form to $\Lambda^2 \Rb^{n+1}$ defined by
\be
\me(e_i \wedge e_j, e_m \wedge e_n ) = \me(e_i, e_m) \me(e_j, e_n)
-\me(e_i, e_n) \me(e_j, e_m)~. \lb{j15}
\ee
After some algebra we get the following decomposition
\be
\me(d\bq ~{\buildrel \otimes \over ,}~ d\bq)
= d q \otimes d q + \frac{1}{q^2}
\me(\bq \wedge d\bq ~{\buildrel \otimes \over ,}~ \bq \wedge d\bq)~, \lb{j16}
\ee
where
\bea
&& q^2 = \me(\bq, \bq) = \me_{ij}q^i q^j~, \nn\\
&& q = \sqrt{\vert q^2\vert }~. \lb{j17}
\eea

\bigskip

Consider now the free motion
on $\Rb^{n+1}$ described by the Lagrangian
\be
\L_{\me} = \frac{1}{2} \me(d\bq ~{\buildrel \otimes \over ,}~ d\bq)(\Ga, \Ga)
= \frac{1}{2}\me_{ij} \td{q}^i\td{q}^j~. \lb{j18}
\ee
It gives
\be
\om_{\L_{\me}} = \me_{ij}dq^i \wedge d\td{q}^j~, \lb{j19}
\ee
\bea
E_{\L_{\me}} & = & \frac{1}{2}\me_{ij} \td{q}^i\td{q}^j \nn \\
&=& \frac{1}{2} \td{q}^2 + \frac{1}{2}
\frac{\me(\bq \wedge \tbq , \bq \wedge \tbq)}{\me(\bq,\bq)} \nn \\
&=& E_{\me}^{rad} + E_{\me}^{ang}~. \lb{j20}
\eea

Consider now the vector field $\Ga_{ang}$  defined by
\be
i_{\Ga_{ang}} \om_{\L_{\me}} = d E_{\me}^{ang}~. \lb{j21}
\ee
We find
\bea
\Ga_{ang} &=& \frac{1}{\me(\bq,\bq)}
\left[\left ( \me(\bq,\bq) \td{q}^j - \me(\bq,\tbq) q^j \right)
\pd{},{q^j} +
\left ( \me(\bq,\tbq) \td{q}^j - \me(\tbq,\tbq) q^j \right)
\pd{},{\td{q}^j} \right] \nn \\
 &+& \frac{1}{\me(\bq,\bq)}
\me(\bq \wedge \tbq , \bq \wedge \tbq) q^j \pd{},{\td{q}^j}
\nn \\
&=& \frac{1}{\me(\bq,\bq)} \Ga_{\me} + \frac{1}{\me(\bq,\bq)}
\me(\bq \wedge \tbq , \bq \wedge \tbq) q^j \pd{},{\td{q}^j}
{}~, \lb{j22}
\eea

\noindent
where $\Ga_{\me}$ is the same as in \rf{j6}.
Therefore, the dynamical system associated with the angular part of
the energy does not coincide with our previous dynamics $\Ga_{\me}$
and moreover
is quite cumbersome. Let us consider however the projection
onto the hyperboloid $\Hb^n = \{\me(\bq,\bq) = 1\}$,
\bea
&& \pi : \Rb^{n+1} \lra \Hb^n~,~~~\bq \mapsto \wt{\bq} =
\frac{\bq}{\sqrt{ \vert \me(\bq,\bq)\vert}},\nn \\
&& T\pi : T\Rb^{n+1} \lra T\Hb^n~,~~~
(\bq,\tbq) \mapsto (\wt{\bq}, \td{\wt{\bq}})~. \lb{j23}
\eea

\noindent
Then $T\pi(q^j \pd{},{\td{q}^j}) = 0$, so that $\Ga_{ang}$ and
$\Ga_{\me}$ project onto the same vector field on $T\Hb^{n}$, i.e.
both give the same geodesical motion on $\Hb^n$.

If we consider these examples in the framework of the unfolding, we
see here that our geodesical motion can be unfolded in two different
ways, one is a simple system, the other one is the angular part of
the free motion.

\bigskip

\bigskip

\sect{Reduction procedure in algebraic terms}\label{se:red}
To illustrate the reduction procedure from the dual view point, we
shall start with the ring $\F$ of smooth functions on a manifold
$M$.

Let us consider again the projection
$\pi_\Si : \F \to \F_\Si$ in
\rf{redAA}. Clearly $ker \pi_\Si$ is an ideal in $\F$ and consists
exactly of functions vanishing on $\F_\Si$, and we have a
corresponding sequence of associative and commutative algebrae
$ 0 \ra ker\pi_\Si \ra \F \ra \F_\Si \ra 0$. The idea is then
that a starting point for an algebraic approach  could be a
sequence of associative, commutative algebrae
\be
 0 \lra \I {\buildrel i \over \lra}~ \F
{\buildrel \pi \over \lra}~ \Q_\I \lra 0~, \lb{redAH}
\ee
where $\I$ is an ideal (with respect to the associative structure) of
$\F$ and $\Q_{\I} = \F / \I$.
Both $i$ and $\pi$ are algebra homomorphisms, e.g.
$\pi(f_1\cdot f_2) =  \pi(f_1) \cdot \pi (f_2)~,
\pi(f_1+f_2)= \pi (f_1)+\pi(f_2)$~.

Any derivation $X$ on $\F$ will define a derivation on $\Q_\I$
if and only if
$X \cdot {\cal I}\subset {\cal I}$, so that
the algebra $\Q_\I$ of equivalence classes is taken into itself. We shall
indicate by $\X(\I)$ the Lie subalgebra of these derivations
\be
\X(\I) = \{X \in Der\F \vdash X \cdot \I \subset \I \}~. \lb{redAI}
\ee

A sequence is compatible with a dynamical vector field $\Ga$ if
$\Ga \in \X(\I)$.
Such a sequence replaces the choice of an invariant submanifold
$\Si$. The analog of the equivalence relation is provided by any
subalgebra of $\Q_\I$ which is invariant under the action of $\Ga$.
The restriction of $\Ga$ to any such subalgebra represents a reduced
dynamics.

Having recalled our algebraic framework for reduction we are now
ready to add additional structures, namely Poisson brackets. As any
symplectic structure is associated with a (non-degenerate)
Poisson bracket we shall not consider symplectic structures
separately.

\bigskip

\subsect{Poisson reduction}
Let us suppose that the algebra $\F$ in \rf{redAH} is a Poisson
algebra. We recall that this means a
Lie algebra structure $\{~, ~\}$ on $\F$ with the
additional requirement that
the map $f \longmapsto X_f$, defined by $X_f\cdot g :=\{f,g\}$,
is a Lie algebra homomorphism from $\F$ into $Der\F$.
We shall first study conditions under which we can reduce also the
Poisson structrure.

We call {\it reduction} of this Poisson structure any quotient
Poisson structure of a Poisson subalgebra $\F' $ of $ \F$
by a Poisson ideal (i.e. an ideal with respect to both the
structures, associative and Lie).

We describe standard ways to obtain reductions.
Let us consider a ``submanifold'', namely an associative ideal $\I$
in $\F$.
If $\I$ is a Poisson ideal we can pass to the quotient. In most
cases, however, this is not true and we must look for a Poisson
subalgebra $\F'$ in $\F$ such that $\F' \cap \I$ is a Poisson ideal.
The algebra $\F'$ can be chosen to be the normalizer
$\N_\I$ of $\I$ under Poisson bracket,
\be
\N_\I =: \{ f \in \F \vdash \{f, \I \} \subset \I \}~.
\ee
It is easy to see that $\N_\I$ is a Poisson algebra [Gr].
Elements in $\N_\I$ are generating functions of Hamiltonian
derivations $X_f$ which project to derivations of
$\Q_\I = \N_{\I} / \I$.  We select a
Poisson subalgebra in $\I$ by intersecting the latter with $\N_\I$.

\bigskip

\noindent
The following statement is true:

The intersection
\be
\I' = \I \cap \N_\I ~,
\ee
is a Poisson ideal in $\N_\I$.

\proof
$\I'$ is already a Lie ideal. From the properties of the Poisson
brackets: $\{\F\I', \I\} \subset \F\{\N_{\I}, \I\} + \I\{\F, \I\}
\subset \I$, so that $\F\I' \subset \N_{\I}$.

\bigskip

\noindent
As an obvious consequence we have:

The space $\Q'_\I =: \N_{\I} / \I'$ is a Poisson algebra such that
the following is an exact  sequence of Poisson algebrae
\be
 0 \lra \I' {\buildrel i \over \lra}~ \N_{\I}
{\buildrel \pi \over \lra}~ \Q_{\I'} \lra 0~. \lb{redAO}
\ee

Given any two elements
(equivalence classes) $ [f], [g] \in \Q_\I' $, their Poisson brackets is
given by
\be
\{ [f], [g]\}_{\Q_I'} = [\{f, g \}_{\N_\I}] ~. \lb{redAP}
\ee

\bigskip

\noindent {\bf Remark.}
In general $\Q_\I$ will be different from $\Q_\I'$.

\bsk

\noindent {\bf Remark.}
Let $\I$ be the ideal $\F = C^{\infty}(M)$ consisting of
smooth functions vanishing on a closed embedded submanifold $\Sigma$.
The normalizer $\N(\I)$ consists of all functions whose Hamiltonian
vector fields are tangent to $\Si$. The quotient algebra $\Q_{\I}$
consists of smooth functions on $\Sigma$.

\bigskip

The second, more geometric method of reducing Poisson structures is
to determine the subalgebra $\F'$ by means of a distribution.
\bsk

\noindent {\bf Definition }
By {\it distribution} on $\F$  we shall mean any subset  $\D$ of the
algebra $Der\F$.  We say that $\D$ is  {\it integrable} if it is a
subalgebra of $Der\F$.

\bsk

Having a distribution $\D \subset Der\F$, define the associative
algebra
\be
\F_{\D} = \{f \in \F \vdash \D \cdot f = 0\}~.
\ee
Note that $\F_{\D} = \F_{\hat{\D}}$, where $\hat{\D}$ is the
integrable distribution generated by $\D$ so we can start with
integrable distributions.

We say that $\D$ is compatible with the Poisson bracket
$\{\cdot, \cdot\}$ if
$\F_{\D}$ is a Poisson subalgebra.

\bigskip

We shall give now an example of reduction associated with a Poisson
bracket that arises as the limit of a quantum group structure on
$SU_q(2)$.

\bigskip

\subsubsect{Example: Lie-Poisson structure for SU(2)}
Consider a quadratic Poisson structure on $\Rb^4$ given by
\be
\begin{array}{lll}
\{y_1, y_2\} = 0 & \{y_1, y_3\} = y_1 y_4
& \{y_1, y_4\} = - y_1 y_3 \\
\{y_2, y_3\} = y_2 y_4 & \{y_2, y_4\} = - y_2 y_3
& \{y_3, y_4\} = y_1^2 + y_2^2~. \lb{glmv1}
\end{array}
\ee
It is easy to show that
$y_1^2 + y_2^2 + y_3^2 + y_4^2 $ is a Casimir function for the
bracket. Therefore,
the latter can be reduced to a bracket on the unit sphere
$S^3$ .
By identifying this sphere with the group $SU(2)$ we get what is
known as a Lie-Poisson structure on $SU(2)$.
We identify the latter with $S^3$ via the map
\be
(y_1, y_2, y_3, y_4) \mapsto s
= y_4 {\bf 1} + i(\vec{y} \cdot \vec{\sigma}) =
\pmatrix{y_4 + iy_3 & -y_2 + iy_1 \cr y_2 + iy_1 & y_4 - iy_3 \cr }~,
\lb{glmv2}
\ee
with $\sigma_i~, i=1,2,3~,$ the three Pauli matrices.

Consider now the compatible distribution $\D$ generated by the vector
field
\be
X = - y_2 \pd{},{y_1} + y_1 \pd{},{y_2} + y_4 \pd{},{y_3}
- y_3 \pd{},{y_4}~. \lb{glmv3}
\ee
The reduced algebra $\F_D$  can be regarded as the algebra of
functions of variables
\be
u = - y_1^2 - y_2^2 + y_3^2 + y_4^2~,~~~
v = 2 (y_1 y_3 + y_2 y_4)~,~~~
z = 2 (y_1 y_4 - y_2 y_3)~, \lb{glmv4}
\ee
with brackets
\bea
&&\{v, u\} = 2(1 - u)z~ \nn \\
&&\{u, z\} = 2(1 - u)v~ \nn \\
&&\{z, v\} = 2(1 - u)u~ \lb{glmv5}
\eea
One finds that $u^2 + v^2 + z^2 = 1$ so that the reduced space of
$SU(2)$ is the
unit sphere $S^2$ and the reduced bracket is singular at the
north pole $(u = 1, v = z = 0)$. The stereographic
projection from the north pole maps this structure onto the standard
one on $\Rb^2$.

\bigskip

\bigskip

Usually, it is more convenient, for computational purposes, to
mix algebraic and geometrical notions, i.e. we use an
ideal (submanifold) and a distribution.

As an example we consider the Poisson reduction of Marsden and Ratiu
[MR].

\bigskip

\subsubsect{Example}
Let us suppose we are given a Poisson manifold $(P, \{\cdot, \cdot\})$
and a submanifold $\Si$ of $P$. Suppose there is a
subbundle $E \subset TP \vert_{\Si}$, with the following properties
\begin{description}
\item[1.] $E \cap T\Si$~ is an integrable subbundle of $T\Si$, so defining
a foliation $\Phi$ on $\Si$;
\item[2.] the foliation $\Phi$ is regular, so that the space of leaves
$\Si / \Phi$ is a manifold and the projection
$\pi : \Si \rightarrow \Si /\Phi$ is a submersion;
\item[3.] the bundle $E$ leaves the Poisson structure invariant in
       the sense that for any two functions $F,~G$
       whose differentials $dF,~dG$ vanish on $E$,
       the Poisson bracket $\{F, G \}$ has differential vanishing
       on $E$;
\end{description}
Define $\F'$ to be the algebra of functions on $P$ whose differential
vanishes on
$E$. Then, condition 3.  assures that $\F'$ is a Poisson
subalgebra of $\F = C^\infty(P)$. If $\I$ is the ideal of functions
vanishing on $\Si$, conditions 1. and 2. assure that $\F' / \I
\cap \F'$ can be thought of as the algebra of smooth functions on
$\Si / \Phi$. To have a Poisson structure on
$\F' / \I \cap \F'$ we must have that $\I \cap \F'$ is a Poisson ideal in
$\F'$, which is clearly equivalent to the fact that Hamiltonian
vector fields associated with $\F'$ when restricted to $\Si$
belong to $T\Si \oplus E$. This is just the last assumption of
Marsden and Ratiu.

\bigskip

The coming example is the typical situation we face when dealing with
the constraint formalism in the Dirac-Bergmann approach.
It should be noticed however that in our approach a dynamics is
already given, it is not to be determined like in the constraint
formalism.

\bigskip

\subsubsect{Example}
Let us suppose we are on a Poisson manifold $(M, \{\cdot, \cdot\})$. We take
$\F \equiv \func$. Given $k$ functions $f_j \in \func~ j \in
\{1, \cdots k\}$ we take the
ideal $\I = \I_a$ generated by
\be
\{f_i - a_i ~\vdash~ a_i \in \Rb,~~ i = 1, \cdots, k \}~.
\ee

Suppose we have any com\-pa\-ti\-ble dis\-tri\-bu\-tion $\D$
ge\-ne\-ra\-ted by vec\-tor
fi\-elds $\{X_i~,~i = 1, \cdots, n \}$. If the $X_i$'s are Hamiltonian
vector fields, then the distribution is automatically compatible.
Put now
\bea
&&\F' = \F_{\D} \cap \N(\I_{a})~, \nn  \\
&&\I' = \I_{a} \cap \F'~. \lb{vienna1}
\eea
It is easy to see that $\F'$ is a Poisson algebra and that $\I'$ is
a Poisson ideal in $\F'$, so that we can reduce the Poisson bracket
to $\F' / \I'$.

Having an $Ad^*$-equivariant momentum map $f = (f_1, \cdots, f_k) : M
\ra \fkg^*$ associated with a Hamiltonian action of the group $G$,
we can take $\D$ generated by the Hamiltonian vector fields
$X_{f_j}$. Then $\F_{\D} \subset \N(\I_{a})$ and we get the classical
reduction with $\F' / \I' = \F_{\D} / \I_a$ interpreted as the algebra
of functions on the space of orbits $\Si_a / G_a$.

\bigskip

\subsubsect{Example}
Consider the cotangent bundle $T^*G$ of a group $G$ with the
canonical Poisson structure. Let $\{x_i\}$ be a basis of the Lie
algebra $\fkg$ and $\{e^i\}$ the dual basis of $\fkg^*$.
Notice that $\{x_i\}$ can
be regarded as a coordinate system on $\fkg^*$. The bundle $T^*G$ can
be naturally considered as a Lie group. In `body' coordinates $T^*G
\equiv G \times \fkg^*$, and the multiplication has the form
\be
(h_1, \mu_1)\cdot(h_2, \mu_2) = (h_1h_2, \mu_1 + Ad^*_{h_1}(\mu_2))~.
\lb{vienna2}
\ee
The Lie algebra of this group can be identified with the semidirect
product $\fkg \oplus \fkg ^*$, with $\fkg ^*$ thought of as a
commutative algebra.

Let $\{X_i, \th^i\}$ be the corresponding left, and $\{\wt{X}_i,
\wth^i\}$
the corresponding right invariant vector fields on $T^*G$. In term of
them the symplectic Poisson structure on $T^*G$ can be written as
\be
\La = \frac{1}{2} ( X_i \wedge \th^i + \wt{X}_i \wedge \wth^i)~. \lb{vienna3}
\ee
In body coordinates $\{h, x_i\}$ for $T^*G$,
\bea
&& \th^i = \pd{},{x_i}~,~~~X_i = X_i^{G} - c_{ij}^k x_k \pd{},{x_j}~,  \\
&& \wth^i = Ad^*_h(\pd{},{x_i})~,~~~\wt{X}_i = \wt{X}_i^{G}~,  \lb{vienna4}
\eea
where $X_i^{G}$ and $\wt{X}_i^{G}$ are the corresponding left and
right invariant vector fields on $G$.
Therefore,
\be
\La = \frac{1}{2} ( X_i^{G} \wedge \pd{},{x_i}
+ \wt{X}_i^{G} \wedge \wth^i
+ c_{ij}^k x_k \pd{},{x_i}\wedge\pd{},{x_j})~. \lb{vienna4a}
\ee

The algebra of functions $\funcc$ can be reduced
to the $\wt{X}_i$-invariant functions $\F'$ using the
projection $\pi : T^*G \equiv G \times \fkg^* \ra \fkg^*$. These
functions can be identified with smooth functions on $\fkg^*$
equipped with the Konstant-Kirillov-Souriou Poisson structure
\be
\La' = \frac{1}{2}
c_{ij}^k x_k \pd{},{x_i}\wedge\pd{},{x_j}~. \lb{vienna5}
\ee

By reducing to the $X_i$-invariant functions, we get the opposite
Poisson structure on $\fkg^*$.

A different way to get the previous structures on $\fkg^*$ is to
consider the left and right invariant momentum maps
\be
\P^{L}~, ~\P^{R} : T^*G \lra \fkg^*~, \lb{vienna6}
\ee
whose components have Poisson brackets
\bea
&& \{P^{L}_i, P^{L}_j \} =- c_{ij}^k P^{L}_k~, \nn \\
&& \{P^{R}_i, P^{R}_j \} =  c_{ij}^k P^{R}_k~, \nn \\
&& \{P^{L}_i, P^{R}_j \} = 0~. \lb{vienna7}
\eea
The algebra $\F''$ of polynomials in $(P^{L}_i, P^{R}_j)$ is a Poisson
algebra. The ideals $\I^{L}$  and $\I^{R}$ generated by left and
right momenta are Poisson ideals, so we can pass to the quotients
$\F''/ \I^{L}$ and $\F''/\I^{R}$ getting as a result the KKS
Poisson structures on $\fkg^*$.

\bigskip

\bigskip

\sect{An example of non-commutative reduction}
Let us see finally, how our algebraic reduction procedure fits into the
non-commutative setting, i. e. to quantum spaces and, after passing to
the semi--classical limit, gives us a ``usual" Poisson reduction.
\smallskip
Roughly speaking, we obtain quantum spaces deforming corresponding
``dual" algebraic objects, as for example the  commutative  algebrae
of a given class functions  on a space  (cf.  Gel'fand-Najmark
functor) can be deformed into noncommutative ones. This class  may
be arbitrarily chosen; Woronowicz [Wo]   prefers   to   work
with $C^*$-algebrae  what  is  more  difficult,  but  fruitful,
while Drinfel'd  works   with   purely   algebraic
version only.
\smallskip
To make the whole thing transparent,  let  us  consider  the  very
classical example   of   Woronowicz [Wo] for   the   group
$SU(2)$  which topologically is a  three  dimensional  sphere.
The *-algebra $\A$ generated by matrix elements
is dense in $C(SU(2))$  and  can  be  characterized  as  the
``maximal" unital commutative *-algebra $\A$ generated by elements
$\alpha ,\nu $  and satisfying  $\alpha  ^*\alpha  +\nu
^*\nu = I$. Dropping the assumption about commutativity,  Woronowicz
proposed  to  consider  the  algebra  $\A_q$  as  the  unital
non-commutative *-algebra generated by $\alpha ,\nu$ satisfying
$\alpha  ^*\alpha  +\nu   ^*\nu   =I$   and    additionally    the
commutation relations
\be
\begin{array}{ll}
\alpha \alpha ^*-\alpha ^*\alpha = (2q-q^2)\nu ^*\nu~,
&\nu ^*\nu -\nu \nu ^*=0~, \\
\nu \alpha - \alpha  \nu  = q\nu  \alpha~, &
\nu ^*\alpha -\alpha \nu ^*=q\nu ^*\alpha. \lb{non1}
\end{array}
\ee
It is clear  that  for  $q=0$  we  get  the  previous  commutative
algebra $\A$, so $\A_q$ is a one-parameter deformation of
$\A=\A_0$. The crucial point here is that the algebrae
$\A_q$ do not ``collapse" for $1>q\ge 0$, i.e. the  basis  remains  the
same (for example $\{ {\nu ^*}^m\nu ^n\alpha  ^k,\  {\nu  ^*}^m\nu
^n{\alpha ^*}^l: k,m,n=0,1,2,...\ ~{\rm and}~ \ l=1,2,...\}$ is a basis  in
$\A_q$ for all $1>q\ge 0$), and we may consider the algebrae
$\A_q$  not  as  different  objects,   but   as   different
multiplications $\circ _q$ on the same object, namely the  space
$\A$. In this way we get the so called {\it  formal  deformations}  of
$\A$, since the $\circ _q$-product of elements  $u,v\in
\A$ reads
\be
u\circ _qv=uv+\sum ^\infty_{n=1}q^nP_n(u,v)~. \lb{non2}
\ee
For example, using the identification of $\A_q$  with
$\A$ via the basis as above, we get
\be
\alpha \circ _q\nu = \nu \circ _q  \alpha  -(\nu  \circ  _q\alpha
-\alpha \circ _q\nu )= \nu \alpha -q\nu  \alpha
\ee
and
\be\alpha \circ _q\alpha ^*=\alpha ^*\circ _q\alpha+(\alpha  \circ
_q\alpha   ^*-\alpha   ^*\circ    _q\alpha)=(I-\nu    ^*\nu)+2q\nu
^*\nu-q^2\nu ^*\nu~.
\ee

Since the commutator bracket $[u, v]_q=u\circ _qv-v\circ _qu$,  as
for any associative algebra, is a biderivation and satisfies the
Jacobi identity, we get easily that
\be
\{ u, v\}:=P_1(u, v)-P_1(v, u)
\ee
is a Poisson bracket on the original commutative algebra $\A.$
Hence the first non-trivial term of a  commutator in the deformed
algebra gives us a Poisson structure on the original manifold.
{}From the commutation relations \rf{non1} defining
the Woronowicz' quantum $3$-sphere  we  get  easily  the  corresponding
Poisson bracket which on matrix elements of $SU(2)$ has the form
\be
\begin{array}{ll}
\{ \alpha , \bar \alpha \} = 2 \bar \nu \nu~,
&\{ \nu ,  \bar  \nu \} = 0~, \\
\{ \nu , \alpha \} = \nu \alpha~ ,
&\{ \bar \nu , \alpha  \}  = \bar \nu \alpha~.
\end{array}
\ee
Passing to real functions $\alpha = y_4 + i y_3~,~\nu = y_2 + i y_1$,
we get  a
purely imaginary bracket with the imaginary part being exactly the
Poisson bracket described in \rf{glmv1}.

Consider now the subalgebra $\A_q'$ in $\A_q$ generated by the elements
$u = I - 2 \nu^* \nu = \alpha^* \alpha - \nu^* \nu~, w = 2 \nu^* \alpha$
and $w^*$. One can easly see
that $uu^* + w^*w = I$ and that $\A_q'$ is a formal deformation of
the algebra $\A_0'$ generated by two complex functions $u$ and $w$
satisfying $\vert u \vert^2 + \vert w \vert^2 = 1$,
namely the algebra of polynomials on the two-dimensional sphere
$S^2$. Therefore, the algebra $\A_q'$ can be regarded as a quantum
$2$-sphere. Since any algebra with a commutator bracket is a Poisson
algebra, our quantum sphere carries a quantum Poisson structure.
One easily finds that
\bea
&&[w, u] = (q^2 - 2q)(1 - u)w~, \nn \\
&&[w^*, u] = - (q^2 - 2q)(1 - u)w^*~, \nn \\
&&[w, w^*] = - 2(q^2 - 2q)(1 - u)  + (4q - 6 q^2 + 4q^3 - q^4)
(1 - u)^2~. \lb{non15}
\eea
Passing to the semi-classical limit and real functions, we get the
functions $u$, $v = Re(w)$ and $z = - Im(w)$ with a purely imaginary
bracket. The brackets of the basic variables are found to be
\bea
&&\{v, u\} = 2(1 - u)z~, \nn \\
&&\{u, z\} = 2(1 - u)v~, \nn \\
&&\{z, v\} = 2(1 - u)u~ \lb{non16}
\eea
which is exactly the structure obtained in \rf{glmv5}.
Therefore, we get the same reduction of the Lie-Poisson structure
using the reduction at the quantum (non-commutative) level and then
passing to the semi-classical limit.

\bigskip

Let us consider now the dynamics on $\A_q$ described by the
Hamiltonian
\be
H = \frac{1}{2} u
= \frac{1}{2} (I - 2 \nu^* \nu)
= \frac{1}{2} (\alpha^* \alpha - \nu^* \nu)~. \lb{non17}
\ee
We have
\bea
&& [H, \nu] = 0~, \nn \\
&& [H, \nu^*] = 0~, \nn \\
&& [H, \alpha] = (q^2 - 2q) \nu^*\nu\alpha~, \nn \\
&& [H, \alpha^*] = - (q^2 - 2q) \nu^*\nu\alpha^*~,    \lb{non18}
\eea
so that the dynamics $e^{i t ad_H}$ is of the form
\bea
&& \nu(t) = \nu~, \nn \\
&& \nu^*(t) = \nu^*~, \nn \\
&& \alpha(t) = e^{i t (q^2 - 2q) \nu^*\nu} \alpha~, \nn \\
&& \alpha^*(t) = e^{- i t (q^2 - 2q) \nu^*\nu} \alpha^*~.   \lb{non19}
\eea
Passing to the classical limit, we get the Hamiltonian
\be
H = \frac{1}{2}(y_4^2 + y_3^2 - y_2^2 - y_1^2)~, \lb{non20}
\ee
which, with respect to the Poisson structure on $S^3$ obtained by
reducing the structure \rf{glmv1}, corresponds to the vector field
\be
\Ga = 2(y_1^2 + y_2^2)(y_4 \pd{},{y_3} - y_3 \pd{},{y_4})~. \lb{non21}
\ee
The corresponding trajectories are given by
\bea
&& y_1(t) = y_1(0)~, \nn \\
&& y_2(t) = y_2(0)~, \nn \\
&& y_3(t) = cos(2t(y_1^2 + y_2^2)) y_3(0)
+ sin(2t(y_1^2 + y_2^2)) y_4(0)~, \nn \\
&& y_4(t) = -sin(2t(y_1^2 + y_2^2)) y_3(0)
+ cos(2t(y_1^2 + y_2^2)) y_4(0)~. \lb{non22}
\eea
These trajectories are really the limits of
\rf{non19} when we take the limit $q^2/q \ra 0$, $q/q \ra 1$, since
then $\nu^*\nu = y_1^2 + y_2^2$ and $\alpha = y_4 + i y_3$. It is
easy to see that $\Ga$ commutes with the vector field $X$ given by
\rf{glmv3}.
Therefore we can reduce the system by the distribution generated by
$X$.

On the other hand, the Hamiltonian \rf{non17} is an element of the
reduced algebra $\A_q'$ and we get a reduced dynamics on $\A_q'$
given by
\bea
&& [H, w] = - \frac{1}{2} (q^2 - 2q)(1 - u) w ~, \nn \\
&& [H, w^*] = \frac{1}{2} (q^2 - 2q)(1 - u) w^* ~,\lb{non23}
\eea
The corresponding solutions for the endomorphism
$e^{i t ad_H}$ are
\bea
&& w(t) = e^{- i t \frac{1}{2}(q^2 - 2q)(1 - u)} w~, \nn \\
&& w^*(t) = e^{i t \frac{1}{2}(q^2 - 2q)(1 - u)} w^*~.   \lb{non24}
\eea
Passing to the semiclassical limit we get the corresponding vector
field on $S^2$,
\be
\wt{\Ga} = (1 - u)(z \pd{},{v} - v \pd{},{z})~, \lb{non25}
\ee
which is really the reduction of $\Ga,$ since
\bea
&& \Ga \cdot u = 0~, \nn \\
&& \Ga \cdot v = 2(y_1^2 + y_2^2)(y_4 y_1 - y_3 y_2) = (1 - u)z~,
\nn \\
&& \Ga \cdot z = - 2(y_1^2 + y_2^2)(y_3 y_1 + y_4 y_2) = - (1 - u)v~.
\lb{non26}
\eea

By using the stereographic projection $: S^2 \ra \Rb^2~, (x, y) =
\frac{1}{1-u}(v, z)$, we obtain the following vector field on
$\Rb^2$:
\be
\Ga(x,y) = \frac{2}{x^2 + y^2 + 1}(x\pd{},{y} - y\pd{},{x})~. \lb{non27}
\ee

\bigskip

\bigskip

\sect{Conclusions}
The guiding idea of this paper is that completely integrable systems
should arise from reduction of simple ones. In particular, many of
them arise from reduction of free systems.

Knowing that these systems
admit alternative Lagrangian or Hamiltonian descriptions, we have
considered the behaviour of these additional structures with respect
to the generalized reduction procedure. We have not addressed the
problem of showing how the recursion operator, known to exist for
many completely integrable systems, arises in the present approach;
however this is not difficult to analyze.

Taking advantage of the duality between a manifold $M$ and the ring
$\func$ of functions defined on it,
we have cast our reduction procedure in an algebraic
framework. As an application we have given a simple example of
reduction in non-commutative geometry.
We may benefit of this approach to deal with the reduction of quantum
systems in the operatorial approach. These aspects will be addressed
in future papers.

\bigskip

\bigskip

\bigskip

\noindent {\Large \bf References}

\begin{description}

\item[AM] Abraham R., Marsden J.E.,
{\it Foundations of Mechanics}, Benjamin,  Reading, MA (1978)

\item[Ar] Arnold V.I., {\it Les me\-tho\-des ma\-the\-ma\-ti\-ques
de la Me\-ca\-ni\-que Clas\-si\-que} (Mir, Moscow, 1976).

\item[Gr] Grabowski J., {\it The Lie structure of $C^*$ and Poisson
       algebras}, Studia Matematica {\bf 81}, 259-270 (1985) .

\item[KKS] Kazdan D., Kostant B. and Sternberg S.,
{\it Hamiltonian group action and dynamical systems of the Calogero
type},
Comm. Pure Applied Math. {\bf 31}, 481-507 (1978).

\item[DMSV] De Filippo S., Marmo G., Salerno M., Vilasi G.,
{\it On the Phase Manifold Geometry of Integrable Nonlinear  Field
Theory}, Preprint IFUSA, Salerno (1982), unpublished.
Il N. Cimento {\bf 83B}, (1984) 97.

\item[LMSV] Landi G., Marmo G., Sparano G. and Vilasi G. {\it A generalized
reduction procedure for dynamical systems},
Mod. Phys. Lett {\bf 6A}, 3445-3453 (1991).

\item[LM] Libermann P. and Marle C.M., {\it Symplectic Geometry
and Analytical Mechanics}, Reidel, Dordrecht, 1987.

\item[MFLMR] Morandi G., Ferrario C., Lo Vecchio G.,
Marmo G. and Rubano C.,
{\it The Inverse problem in the Calculus of variations
and the geometry of the
tangent bundle}, Phys. Rep. {\bf 188}, 147-284 (1990).

\item[MM] Man'ko V.I. and Marmo G.,
{\it Generalized Reduction Procedure and Nonlinear
Nonstationary Dynamical Systems}, Mod. Phys. Lett. {\bf A7} (1992)
3411-18.

\item[Mo] Moser J., {\it Various Aspects of Integrable Hamiltonian Systems},
in  Dynamical Systems, Progr. Math. {\bf 8}, Birkh\"{a}user, Boston
(1980).

\item[MSSV] Marmo G., Saletan E.J., Simoni A. and Vitale B. {\it Dynamical
Systems, a Differential Geometric Approach to Symmetry and Reduction} John
Wiley, Chichester UK 1985.

\item[MR] Marsden J., Ratiu T., {\it Reduction of Poisson manifold},
       Lett. Math. Phys. {\bf 11}, (1986) 161-169.

\item[OP] Olshanetsky M.A., Perelomov A.M., {\it Classical Integrable
       Finite Dimentional Systems Related to Lie Agebras},
       Phys. Rep. {\bf 71}, 313 (1981); {\it Quantum Integrable
       Systems Related to Lie Agebras}, Phys. Rep. {\bf 94}, 313 (1983);

\item[ORW] del Olmo M.A., Rodriguez M.A., Winternitz P., {\it Integrable
       Systems Based on SU(p,q) Homogeneous Manifolds}, Preprint
       Universidade de Valladolid, 1992.

\item [Wi] Whittaker E.T., {\it A treatise on the analytical dynamics of
particles and rigid bodies}, Cambridge University Press, London
(1904, 4th edn, 1959).

\item[Wo] Woronowicz S. L., {\it Twisted SU(2)  group.  An  example  of  a
non commutative differential calculus}, Publ.  Res.  Inst.  Math.
Sci. {\bf 23}, (1987) 117-181.

\end{description}

\end{document}